\documentclass[11pt]{article}

\usepackage[parsep]{collref}

\usepackage{hyperref}
\hypersetup{
    colorlinks=true,
    linkcolor=blue,
    filecolor=magenta,      
    urlcolor=blue,
    citecolor=blue,
    linktocpage=true,
    }

\usepackage{latexsym,amsmath,amssymb,theorem,epsfig}
\DeclareFontFamily{OT1}{rsfs10}{}
\DeclareFontShape{OT1}{rsfs10}{m}{n}{ <-> rsfs10 }{}
\DeclareMathAlphabet{\mathscript}{OT1}{rsfs10}{m}{n}

\numberwithin{equation}{section}

\newcommand{\tr}{\text{tr}}

\newcommand{\RR}{{\mathbf{\rR}}}

\newcommand{\com}[2]{[#1,#2]}
\newcommand{\anticom}[2]{\{#1,#2\}}

\def\a{\alpha}

\def\d{\delta}

\def\k{\kappa}
\def\l{\lambda}
\def\m{\mu}

\def\r{\rho}
\def\s{\sigma}
\def\t{\tau}

\def\G{\Gamma}

\def\P{\Pi}

\def \S {{\cal S}}

\def\gsim{ \lower .75ex \hbox{$\sim$} \llap{\raise .27ex \hbox{$>$}} }
\def\lsim{ \lower .75ex \hbox{$\sim$} \llap{\raise .27ex \hbox{$<$}} }
\def\be{\begin{equation}}
\def\ee{\end{equation}}
\def\bea{\begin{eqnarray}}
\def\eea{\end{eqnarray}}

\def \ha {{1 \ov 2}}

\def \del{\partial}
\def \a {\alpha}

\def\ov{\over}
\def \ci {\cite}

\def \foot {\footnote}

\def\la{\label}
\def\foot{\footnote}
\newcommand{\rf}[1]{(\ref{#1})}
\def \OO {{\cal  O}}\def \no {\nonumber}

\def \LL {{\rm L}}
\def \N {{\cal N}}

 \def \hD {\hat \D}
\def \ov {\over}

\def \Ima {\mathrm{Im}}

\def \ed {
\small
\bibliography{biblio2.bib}
\bibliographystyle{JHEP-v2.9}
\end{document}
}

\def \edo {\end{document}}

\def \N {{\cal N}}
 
\def \Tr {{\rm Tr}}
\def \iffa {\iffalse}

\usepackage{tikz}
\usetikzlibrary{decorations.markings}

\makeatletter
\renewcommand*{\@fnsymbol}[1]{\textit{\@alph{#1}}}
\makeatother

\begin{document}
\begin{titlepage}
\vspace{-15cm}
\vspace{-8cm}
\title{\vspace{-3cm}
  \hfill{\small Imperial-TP-FS-2024-01  }  
  \\ \vspace{1cm} 
   { 
   Scattering on the supermembrane }
  \\[1em] }

\author{\large Fiona K. Seibold\thanks{f.seibold21@imperial.ac.uk} \, 
    and   Arkady A. Tseytlin\footnote{Also at  the Institute  for Theoretical and Mathematical Physics    (ITMP)  of  MSU
      and  Lebedev
    Institute.}  
      \\[1.3em]
    {   Blackett Laboratory, Imperial College, London SW7 2AZ, U.K. }
}
  \maketitle
{\small
\begin{abstract}
We compute the one-loop $2 \to 2$ scattering  amplitude of  massless    scalars on 
the world volume of  an  infinite  $D=11$  supermembrane   quantized in the  static gauge. 
The resulting expression  is manifestly  finite and  turns out to be 
  much simpler than in the bosonic membrane case  in 
arXiv:2308.12189  being  proportional to the tree-level scattering amplitude. 
We also  consider the case of   $\mathbb R^{1,1} \times S^1$  membrane   with  one  dimension 
  compactified on  a circle of radius R  and  demonstrate 
how the supermembrane scattering  amplitude  reduces to the one on an  infinite
  $D=10$  Green-Schwarz  superstring   in the limit of R$\to 0$.   
\end{abstract}
}

\end{titlepage}

\def \iffa  {\iffalse}
\def \RR {{\mathbb R}}
\def \R {{\rm R}}
\def \s {\sigma} \def \t {\tau} 
\def \G {\Gamma} \def \four {{1\ov 4}}
\def \CP  {{\rm CP}}\def \gs {g_s}
\def \hD   {\hat{D}}
\def \JJ {{\rm J}}
\def \te {\textstyle}
\def \zz  {^{(0)}} 
\def \ve {\varepsilon}
\def \zo {^{(1)}}
 \def \rR  {{\rm R}} 
 \def \eps {\epsilon}
 \def \rS  {{\rm S}} \def \rT {{\rm T}}
\def \P   {{\rm P}}
 \def \bbeta {{\rm r}} 
\def \no {\nonumber}
\def \TTT  {{\cal T}} 
\def \S {{\rm S}}  \def \rT  {{\cal T}} 
\def \dDelta {{\hat \Delta}}
\def \ss {\tau}
\def \tn {\lambda} 
\def \hD {\hat D}
\def \half {\tfrac{1}{2}}
\tableofcontents
%
\def \ba {\begin{align}}
\def \ea {\end{align}}
\def \RR  {{\mathbb R}}
\def \rr {{\rm R}}  \def \bQ {{\bar Q}}
\def \nf {{\rm n}_{_{\rm F}}}
\def \hT  {\hat T}

 \def \bD {{\rm  b}}  
   \def \rD {{\rm D}} 
\def \LL {{\rm L}}
\def \vt {\vartheta}
\def \ln {{\rm log\,}}
\def \tad {{\rm tad}}

\def \com {}

\section{Introduction  and summary}

Despite its  non-linearity and formal  non-renormalizability,  
the 3d   action  of  the  supermembrane 
propagating in    special  11d supergravity backgrounds~\ci{Bergshoeff:1987cm,Bergshoeff:1987qx} (see~\ci{deWit:1988wri,Duff:1996zn,Nicolai:1998ic} for reviews) 
may    correspond to   a consistent quantum  field theory 
  when expanded  near a non-degenerate world volume solution
(cf.~\ci{Duff:1987cs,Mezincescu:1987kj,Forste:1999yj,Becker:1995kb,Harvey:1999as}).  
An ample evidence  that quantum  supermembrane (M2 brane)  theory in  the maximally   supersymmetric 
AdS$_4 \times S^7$ or  AdS$_7 \times S^4$  backgrounds~\ci{Bergshoeff:1988uc,Duff:1989ez,deWit:1998yu} 
is well  defined in the semiclassical expansion  was recently provided in the context of the 
AdS/CFT duality~\ci{Drukker:2020swu,Giombi:2023vzu,Beccaria:2023ujc,Beccaria:2023sph,Drukker:2023jxp,Drukker:2023bip}.
The 3-dimensional   M2  brane theory  has no   non-trivial    UV  divergences at one loop, and  its  extended supersymmetry 
 and possibly  other  hidden  symmetries   may insure also  the absence  of     higher-loop  divergences.  This 
   is, in fact,   required for the expected matching with  localization and defect  anomaly  predictions  on the dual CFT  side
 (cf.~\ci{Drukker:2020swu,Giombi:2023vzu,Beccaria:2023ujc,Beccaria:2023sph}). 
 
In an attempt to uncover some  hidden simplicity   of the 11d   M2 brane  theory, one may start with the flat target-space case and study  the  S-matrix of  massless  transverse  excitations propagating on an infinite  membrane.
The analogous S-matrix of massless excitations on the long (super)string has  a very special (elastic, pure phase) form reflecting  the  integrability of the world-sheet theory~\ci{Dubovsky:2012sh,Dubovsky:2012wk,Cooper:2014noa,Mohsen:2016lch}.
While there is no direct analog of integrability in the 3d   case,  one  may expect that the S-matrix on  the supermembrane may  
have  some  unique  features compared  to the  bosonic  membrane case    considered in~\ci{Seibold:2023zkz}.  The aim of the present   work is to demonstrate that this is indeed the case.

We shall  consider  the  scattering of massless ``transverse" $\hD= D-3=8$  scalars $X^I$ 
  propagating on $\mathbb R^{1,2}$  world volume of an infinite membrane  vacuum. 
  The static-gauge   action  has the following symbolic form $ T_2\int d^3\s [(\del X)^2 + (\del X)^4 + ...  + \bar \theta\del \theta  + (\del X)^2 \, \bar \theta\del \theta + ...]$ where $\theta$ are the  ``physical"  components of the  fermions after  $\kappa$-symmetry gauge fixing  and dots stand for terms  with more derivatives.  Here $T_2= {1\ov (2\pi)^2 \ell_{pl}^3}$ is the  M2 brane  tension. 
  The  amplitude 
   of   the $2 \rightarrow 2$ scattering  between two incoming bosons 
   with $SO(\hD)$ indices $I_1$,$I_2$ and 3d momenta $p_1$,$p_2$, and two outgoing bosons with  $I_3$,$I_4$ and $p_3$,$p_4$   has the following general structure~\foot{For notational simplicity we will  suppress the  
   indices $I_1,I_2,I_3,I_4$ on $\mathcal M$. We will also omit the overall factor $T_2^{-1} $ present already in the tree-level amplitude
   and the momentum  conservation delta-function.}
   \begin{align}  \la{1}
& \mathcal M=A \, \delta^{I_1 I_2} \delta^{I_3 I_4} + B \, \delta^{I_1 I_3} \delta^{I_2 I_4} + C \, \delta^{I_1 I_4} \delta^{I_2 I_3}~, \\
&\   \la{11} B = A(s,t,u) \Big|_{s \leftrightarrow t}~, \qquad \qquad  C= A(s,t,u)\Big|_{s \leftrightarrow u}~, \qquad \ \  s+t+u=0 \ , 
\end{align}
where $A$ (``annihilation"), $B$ (``transmission")  and $C$ (``reflection")   amplitudes 
 are   functions of the Mandelstam variables $s,t,u$  
  related by  crossing  symmetry.
 Using standard  perturbation theory  one finds that 
\begin{align}\la{2} 
&\mathcal M =  \mathcal M^{(0)} + T_2^{-1} \mathcal M^{(1)} + T_2^{-2} \mathcal M^{(2)}+    \dots~,\\
\la{3}
&\mathcal M^{(0)} = -\tfrac{1}{2} \left( t u \, \delta^{I_1 I_2} \delta^{I_3 I_4}  +  s u \, \delta^{I_1 I_3} \delta^{I_2 I_4} + s t \, \delta^{I_1 I_4} \delta^{I_2 I_3} \right)~, 
\end{align}
where  $\mathcal M^{(n)} $   stand for  $n$-loop corrections.  The tree-level   amplitude  $ \mathcal M^{(0)} $ 
is  the same as in the bosonic  membrane  case.\foot{ {\com  The  double softness  of this amplitude 
 may be  related to the fact that 
the  fields in the  Lagrangian may be interpreted as 
the Nambu-Goldstone particles   corresponding to spontaneous breaking of the Poincare symmetry 
 (and global supersymmetry) of the original  supermembrane action by the choice of the static-gauge
  vacuum (cf.   \ci{Heydeman:2017yww,Goon:2020myi}  and refs. there).}}

{\com  As we shall  find below,  the 1-loop  correction  to the amplitude   has  the same kinematical structure as the tree-level one, i.e. $A$ in \rf{1}   is  again proportional to $tu$, etc.  If we conjecture that (due to  underlying supersymmetry) 
the same  pattern will apply also at higher loop orders  then,   }
on dimensional grounds, 
 the amplitudes $A,B,C$  in~\rf{1}   may be written as functions of dimensionless ratios, e.g., 
\begin{equation} \la{4} 
A=  -\half  tu \Big[ 1 +  T_2^{-1} s^{3/2}  f^{(1)} ( {s\ov t}) +  T_2^{-2} s^{3}  f^{(2)} ( {s\ov t})  + ...\Big] 
\ . 
\end{equation}
Here we assumed  that there is  no extra  dimensional scale (UV cutoff)   appearing    in   loop   corrections
{\com (as we are considering perturbation theory  in $ T_2^{-1}$,  no  non-analytic terms like 
$\log ( s  T_2^{-2/3} )$    may appear).}
While this is manifestly so at the 1-loop order that we will focus on below,  to check that  this  
 happens  also at the 2-loop order  is an  important     open problem.~\foot{In general,   if  log  UV  divergences  will not automatically cancel at $\ell \geq 2$   loop orders  this will require to add  local counterterms  to cancel them order by order in  loop  expansion. One may conjecture that the  structure of these counterterms  will be controlled   by some hidden  symmetry  that  will then define the  quantum M2  brane   theory  uniquely.}

A priori, one may expect  $ f^{(1)} ( {s\ov t}) $ in~\rf{4}   to be a  non-trivial   function.  
Indeed, in the case of the bosonic membrane in 
$D= \hD +3$   dimensions  one finds the following expression for the 1-loop amplitude in~\rf{2}~\ci{Seibold:2023zkz}
\begin{align}
&\mathcal M_B^{(1)}  = \tfrac{1}{256} \delta^{I_1 I_2} \delta^{I_3 I_4}   
\Big[(-s)^{3/2}   \big[ \big(\tfrac{3\hD}{32}-1\big) s^2 - \tfrac{\hD}{4} t u  \big] + (-t)^{3/2}  t(3t+2s)  + (-u)^{3/2} u(3u+2 s) \Big]\no  \\
 &\qquad \ + \tfrac{1}{256} \delta^{I_1 I_3} \delta^{I_2 I_4} \Big[ (-t)^{3/2}  \big[ \big(\tfrac{3 \hD}{32}-1\big) t^2 - \tfrac{\hD}{4} s u  \big] +  (-s)^{3/2} s(3s+2t) + (-u)^{3/2} u(3u+2t) \Big]\la{5} \\
 &\qquad \ +\tfrac{1}{256} \delta^{I_1 I_4} \delta^{I_2 I_3}  \Big[ (-u)^{3/2} \big[ \big(\tfrac{3 \hD}{32}-1\big) u^2 - \tfrac{\hD}{4} s t  \big] + (-s)^{3/2} s(3s+2 u) + (-t)^{3/2} t(3t+2 u)  \Big]~.\no 
\end{align}
Compared to the  bosonic string  case where the 1-loop 
amplitude takes  the special form   for the critical  dimension  $26$~\ci{Dubovsky:2012sh}
here there is no similar distinguished   value of $\hD=D-3$. 

Remarkably,  a drastic simplification occurs  when we   add fermions, i.e.~consider the supermembrane 
 in the maximal dimension $D=11$.  
 While   the classical supermembrane action   can be   formally 
 defined  for  several special    target-space dimensions $D=4,5,7,11$~\ci{Achucarro:1987nc}, 
 the   1-loop   scattering amplitude  simplifies  only for  $D=11$.~\foot{While there are arguments  that,  like the $D=10$ superstring,  
   the quantum supermembrane theory may be 
  well-defined  only  in the special dimension    $D=11$~\ci{Bars:1987nr,Marquard:1989rd,Meissner:2022lso}
  we will formally  compute  the  1-loop   scattering amplitude  also for  other classically allowed  values of  
   $D$.}

 As we will  demonstrate  below, combining the bosonic loop  contribution~\rf{5}   with that of  the fermionic loop   gives the total amplitude with $f^{(1)} $  in~\rf{4} being  just a constant. As a result, 
the  $D=11$  supermembrane 1-loop amplitude  is   proportional to the tree-level one in~\rf{3}, i.e. 
\begin{equation}\la{6}
\mathcal M^{(1)} = - \tfrac{1}{64}  \Big[ (-s)^{3/2} + (-t)^{3/2} + (-u)^{3/2} \Big] \Big(tu\, \delta^{I_1 I_2} \delta^{I_3 I_4}   +su\,  \delta^{I_1 I_3} \delta^{I_2 I_4}    + st\,  \delta^{I_1 I_4} \delta^{I_2 I_3}   \Big)~.
\end{equation}
One may  conjecture   that  the same pattern  may apply   also to higher-loop corrections. 
{\com  Indeed, this happens, e.g.,   in the $\N=4$ SYM theory where  the amplitude is proportional to the tree-level one 
as a consequence of maximal supersymmetry.}
{\com Assuming  that  higher loop  corrections to \rf{6}   will continue to be  UV  finite,    we may  make a further 
assumption that they  will depend  just on powers 
of a single   dimensionless  kinematic  parameter $\r$} \foot{\com It would be interesting to check  if this assumption is actually consistent with unitarity. Checking  perturbative unitarity requires computation of  amplitudes involving fermions
(see Appendix \ref{B})  which remains an interesting open problem. }
\begin{align}
&\mathcal M =F( \rho)\,   \mathcal M^{(0)}\,  ,  \qquad \qquad 
F(\r) = 1  + c_1  \rho  + c_2  \rho^2 +    \dots ~,\qquad \ \ \  c_1=\tfrac{1}{32} \ , 
\la{7}\\
&\qquad  \rho\equiv  T_2^{-1}\big[ (-s)^{3/2} + (-t)^{3/2} + (-u)^{3/2} \big]  =   T_2^{-1}  s^{3/2} \big( \sin^3\tfrac{\varphi}{ 2}  +\cos^3\tfrac{\varphi}{2} - i  \big) 
      \ .  \la{8} 
\end{align} 
Here $\varphi$  is the   scattering  angle  in the c.o.m.~frame where 
\begin{equation} \la{9}  
s>0~,\ \qquad     t = -\tfrac{1}{2}s (1-\cos \varphi) <0~, \ \qquad  u=-\tfrac{1}{2} s(1+\cos  \varphi)<0 \ . 
\end{equation}
The   amplitude~\rf{6},~\rf{7} contains an  imaginary part as required by  unitarity. 
Since the theory  has a   dimensional  coupling $T_2$ the  perturbative  amplitudes grow 
with energy. 
One  may hope that the exact function $F( \rho)$  may behave  at large $s$ in a way consistent with  the
 Froissart bound.  For example,  one may  {\com make a further bold conjecture that  1-loop correction exponentiates  so that}\foot{\com  Such  an exponentiation to a   phase   happens in the case of  integrable 2d models like Nambu   string \ci{Dubovsky:2012sh}
 but whether it  is consistent with unitarity in  the present  case  is  far  from    clear. {\com Similar exponentiation is known to happen in  the 4-point amplitude in $\N=4$ SYM theory \ci{Anastasiou:2003kj,Bern:2005iz,Alday:2008yw}  (and 
also in closely related  case of  the 3d ABJM theory \ci{Bianchi:2011aa})  where   it is related  to  supersymmetry but also to dual conformal 
symmetry,   suggesting  that if  it happens  in the present case that may also be an indication of some hidden  symmetry
(we  thank R. Roiban for this remark). } }
 \begin{equation}\la{88} 
  F(\r)= e^{ c_1 \r} = \exp\big[  \tfrac{1}{32} T_2^{-1}  s^{3/2} \big( \sin^3\tfrac{\varphi}{ 2}  +\cos^3\tfrac{\varphi}{2} - i  \big) 
\big]\ . 
\end{equation} 

Following~\ci{Seibold:2023zkz}   we may also consider the   massless boson 
scattering  on the cylindrical  $\RR^{1,1} \times S^1$   membrane   vacuum   with the  circular dimension  having 
    radius $\rr$. 
 Choosing  the 3d  static gauge and Fourier expanding in the compact  world-volume coordinate,  the bosonic membrane action   becomes equivalent to 
 an effective 2d model   containing  the massless  sector represented by the Nambu  string action in the 2d static gauge 
 and  an infinite  tower of massive modes. 
 The full 2d theory   containing both massless  and massive modes is not integrable~\ci{Seibold:2023zkz}   but  one  may still   study the  S-matrix  in the purely  massless (i.e.~string-mode) sector, 
  with the massive modes  propagating only 
 in the loops. 
 
 A similar analysis   can be carried out in  the case  of the supermembrane,   where the corresponding   massless sector will be representing the modes of  the  type IIA Green-Schwarz (GS)  superstring~\ci{Duff:1987bx}. For the 2d massless kinematics one may assume that 
 $t=0, \ u=-s$ so that the tree level amplitude~\rf{3} reduces to the purely elastic $\sim  \delta^{I_1 I_3} \delta^{I_2 I_4} $ part. 
 As we will show below the same   is true also  for the 1-loop  correction
  \begin{align}
&\hat {\mathcal M} =  \hat F (s)\,  \hat {\mathcal M}^{(0)}  \ , \ \ \ \qquad \qquad  
\hat {\mathcal M}^{(0)} =  \tfrac{1}{2}   s^2  \, \delta^{I_1 I_3} \delta^{I_2 I_4} \ , \la{10} \\
&   \hat F (s) = 1  + \tfrac{1}{16 \pi}  T_2^{-1} \rr^{-1}  s \Big(i  -    \tfrac{1 }{\pi } \sum_{n=1}^\infty \big[\bQ_n(\rr^2s)- \bQ_n(-\rr^2s)\big] \Big) + \OO( T_2^{-2}) 
\ , 
\la{111}\\
&\bQ_n(x) \equiv   - \tfrac{2}{ \sqrt{1+\frac{4 n^2}{x}}} \ln   \tfrac{ \sqrt{1+\frac{4 n^2}{x}}-1}{\sqrt{1+\frac{4 n^2}{x}}+1}~.
\la{12}
\end{align} 
This  1-loop amplitude  is again much simpler than  its counterpart in the  case of the bosonic membrane~\ci{Seibold:2023zkz}
(which contains  also  non-zero $A^{(1)}$ and $C^{(1)}$  contributions in~\rf{1}). 

In the decompactification  limit  $\rr \to \infty$  the amplitude~\rf{10}  reduces to the 1-loop amplitude in~\rf{7} in the special  case of 
scattering in a  2-plane (with $\varphi=0$  in~\rf{9}). In the opposite limit
$\rr\to 0$ (with fixed string tension $T_1= 2 \pi \rr T_2$) 
 one recovers the  ``pure-phase''  elastic  scattering  amplitude  in the  $D=10$ GS  string in the static gauge 
 (cf.~\ci{Seibold:2023zkz}). 

\medskip\medskip

The structure of this paper is as follows. 
In section~\ref{s2} we define the supermembrane action and present the first few orders in the perturbative expansion around the flat
$\mathbb R^{1,2}$ membrane  vacuum  in the static gauge  and  the corresponding 
tree level amplitude for the scattering of 4  massless bosons. 
 In section~\ref{s3} we compute the 1-loop  correction to this  amplitude  and demonstrate  its remarkable simplification 
 in the case of the $D=11$ supermembrane. 
 
 In section~\ref{s4}  we consider  the  supermembrane   in  the  $\mathbb R^{1,1}\times S^1$  vacuum 
 where it may be represented as an effective 2d theory containing the massless  superstring sector and an infinite tower  of massive 2d modes. We compute the 1-loop   scattering amplitude for massless 2d  bosons   and  find 
  that it  becomes diagonal (i.e.~with only transmission  $B$ amplitude being non-zero)  in the special 
  case of the  $D=11$  supermembrane.
  We  discuss  the  limits when the radius $\rr$  of the circle  goes to zero  or infinity. 
In the  $\rr\to 0$ limit  one  recovers the  diagonal   scattering amplitude in the  critical GS superstring. 
  In section~\ref{s5} we  discuss   how this conclusion can  reached  by   the direct 
   computation of the   scattering amplitude   from the GS action  expanded near the infinite string vacuum. 
   
   Appendix~\ref{A} summarises our conventions, some identities   and fermionic spectrum of   supermembrane models.
   Appendix~\ref{B} comments on  consistency of the imaginary part of the 1-loop  amplitude with the unitarity of  the S-matrix. 
   Appendix~\ref{C} presents the computation of the contribution of   tadpole diagrams.


\section{The supermembrane action and tree-level amplitude}
\label{s2}

The     supermembrane   action in flat  target space   has the following   action~\ci{Bergshoeff:1987cm}
\begin{align}
&S = S_1 + S_2~,\label{13} \qquad \qquad 
S_1 = -T_2\int d^3 \sigma \, \sqrt{-\det g_{\mu \nu}}  ~,  \\
&S_2 = -T_2\int d^3 \sigma \ \tfrac{i}{2} \epsilon^{\mu \nu \lambda} \bar{\theta} \Gamma_{MN} \partial_\mu \theta \Big( \Pi_\nu^M \Pi_\lambda^N + i \Pi_\nu^M \bar{\theta} \Gamma^N \partial_\lambda \theta - \tfrac{1}{3} \bar{\theta} \Gamma^M \partial_\nu \theta \, \bar{\theta} \Gamma^N \partial_\lambda \theta  \Big) ~,\la{14} \\ 
& g_{\mu \nu} = \eta_{MN}  \Pi^M_\mu \Pi^N_\nu~, \qquad \Pi_\mu^M = \partial_\mu X^M - i \bar{\theta} \Gamma^M \partial_\mu \theta~, \qquad \bar{\theta} = \theta^\dagger \Gamma^0~, \ \ \  \anticom{\Gamma^M}{\Gamma^N} = 2 \eta^{M N} \ . \la{15}
\end{align}
Here  $\sigma^\mu=(\sigma^0,\sigma^1,\sigma^2)$  are the world-volume coordinates  
 while $X^M$ are the target-space coordinates  with $M=0,1,2,\dots,D-1$.~\foot{We assume that
   $\eta_{\mu\nu}=\text{diag}(-1,1,  1), \  \eta_{MN}=\text{diag}(-1,1,..., 1)$   and $\epsilon^{012}=-1$.
   We define  $\Gamma^{M_1 M_2, \dots M_n} = \Gamma^{[M_1}\Gamma^{M_2} \dots \Gamma^{M_n]}$ as $ { 1\ov n!}( \Gamma^{M_1}\Gamma^{M_2} \dots \Gamma^{M_n} + ...)$,  so that 
    $\Gamma^{MN} = \frac{1}{2}(\Gamma^M \Gamma^N - \Gamma^N \Gamma^M)$, etc.}
     To ensure the  reality of the  kinetic term for the fermions we also  assume that 
$
\Gamma^0 (\Gamma^M)^\dagger \Gamma^0 = \Gamma^M~.
$
 $\Pi_\mu^M$ and hence $S_1$ are  invariant under the rigid  target-space supersymmetry ($\ve$ is a constant spinor)
\begin{equation}
\delta X^M =  i \bar{\ve} \Gamma^M \theta~, \qquad \qquad \delta \theta = \ve~. \la{16}
\end{equation}
The WZ  term $S_2$ in~\rf{14}  in  also  supersymmetric (up to a total divergence) if the  following  identity is satisfied
(see e.g.~\ci{deWit:1988wri}) 
\begin{equation}
\bar{\vt}_{[1} \Gamma^M \vt_2\ \bar{\vt}_3 \Gamma_{MN} \vt_{4]}=0~, \la{17}
\end{equation}
where the brackets denote antisymmetrisation over the four arbitrary spinors $\vt_{1,2,3,4}$.
This  holds  for  $D=4,5,7,11$~\ci{Achucarro:1987nc}  with an appropriate choice of 
spinors. 
In the maximal dimension $D=11$ the spinors  should  be  Majorana, i.e.~obeying the  condition $\bar{\theta} = \theta^t \mathcal C$  (thus  having 
 32 independent real components). 
  The charge conjugation matrix $\mathcal C$ satisfies $\mathcal C \Gamma^M \mathcal C^{-t} = (\Gamma^M)^t$, $\mathcal C^t = -\mathcal C$ and $\mathcal C^2=-1$.
   This implies that~\foot{The choice of spinors in other dimensions is 
   discussed in Appendix~\ref{A}. In other dimensions  similar 
   expressions in the action should always be appropriately symmetrised.
}  
$(\bar{\theta} \Gamma^{N_1 \dots N_n} \partial_\mu \theta)^t = - (-1)^{n-1} \partial_\mu \bar{\theta} \Gamma^{N_n \dots N_1} \theta~.$

The relative coefficient  between $S_1$ and $S_2$ in~\rf{13} is 
 chosen so that the action $S$  has also a local fermionic $\k$-symmetry (which is true for all  special dimensions  $D=4,5,7,11$)
\begin{equation} \label{19}
\begin{aligned} 
\delta X^M = i \bar{\theta} \Gamma^M (1+\mathbf{\Gamma}) \kappa~, \qquad \delta \theta = (1+\mathbf{\Gamma}) \kappa~.
\end{aligned}
\end{equation}
Here  $\kappa$ an arbitrary $\sigma$-dependent spinor and
\begin{equation} \label{20}
\mathbf{\Gamma} \equiv  \tfrac{1}{6 \sqrt{-g}} \epsilon^{\mu \nu \lambda} \Pi^M_\mu \Pi^N_\nu \Pi^P_\lambda  \Gamma_M \Gamma_N \Gamma_P~, \qquad\qquad  \mathbf{\Gamma}^2 =1~.
\end{equation}
We will be interested in expanding the action $S$  near the  vacuum  corresponding to infinite $\mathbb R^{1,2}$ membrane. 
In this case  it is natural to choose the static gauge 
\begin{equation} \label{21}
X^\mu =\s^\mu, \qquad\qquad  \mu=0,1,2 \ .
\end{equation}
This leaves $\hD = D-3$ dynamical  bosonic  fields $X^I$ ($ I=3,\dots,D-1$), describing fluctuations in the 
directions transverse to the membrane.~\foot{As usual, these  may  be viewed as the Goldstone bosons associated to the symmetry breaking $ISO(1,D-1) \rightarrow ISO(1,2) \times SO(D-3)$ realised by the membrane vacuum configuration. While  the action in~\rf{13} 
 is invariant under the  global Poincar\'e  group  $ISO(1,D-1)$,   after static gauge fixing it is only manifestly invariant under the the world volume $ISO(1,2)$ symmetry  and $SO(D-3)$ transverse  rotations, with the rest of the 
 original symmetries  is realised non-linearly.}

Expanding  the action in powers of $X^I$ and $\theta$  we get
\begin{align} \label{23}
S  = \, &T_2 \int d^3 \sigma \,\mathcal L~, \qquad \qquad \ \ \mathcal L =  -1 + \mathcal L_2 + \mathcal L_3 + \mathcal L_4 + \dots~,
\\
\mathcal L_2 = \,&-\tfrac{1}{2} \partial_\mu X^I \partial^\mu X^I +i\bar{\theta} \left(1  - \Gamma^\star \right)  \Gamma^\mu \partial_\mu  \theta~, \la{24} \\
\mathcal L_3 = \,&\ i \partial_\mu X^I \bar{\theta}    \Gamma^I \partial^\mu \theta  + {i} \partial_\mu X^I  \bar{\theta} \Gamma^\star \Gamma^I  \Gamma^{\mu\nu}\partial_\nu \theta  ~, \la{25} \\
\mathcal L_4 =\,& -\tfrac{1}{8} \partial_\mu X^I \partial^\mu X^I \partial_\nu X^J \partial^\nu X^J +\frac{1}{4} \partial_\mu X^I \partial_\nu X^I \partial^\nu X^J \partial^\mu X^J\no  \\
& + \tfrac{i}{2} \partial_\mu X^I \partial^\mu X^I \bar{\theta} \Gamma^\nu \partial_\nu \theta - i \partial_\mu X^I \partial^\nu X^I \bar{\theta} \Gamma^\mu  \partial_\nu\theta\no \\
&- \tfrac{i}{2} \partial_\nu X^I \partial_\lambda X^J \epsilon^{\mu \nu \lambda} \bar{\theta}  \Gamma^{IJ} \partial_\mu  \theta + \OO(\theta^4)~.\la{27}
\end{align}
We  used the relations
\begin{equation}
\Gamma^\star \equiv  \Gamma^0 \Gamma^1 \Gamma^2~,\qquad  (\Gamma^\star)^2 = 1 \, , \qquad 
\epsilon^{\mu \nu \lambda} \Gamma_{\nu \lambda} = 2 \Gamma^\star \Gamma^\mu~, \qquad \epsilon^{\mu \nu \lambda} \Gamma_{I \lambda} = \Gamma^\star \Gamma_I \Gamma^{\mu\nu} \ . \la{22}
\end{equation}
We will   consider only the 1-loop 
 scattering of the bosons $X^I$   so will not  need the explicit form of the $\theta^4$ vertices. 
To  get a non-degenerate  kinetic term for the fermions we impose the following $\k$-symmetry gauge 
\begin{align} 
& \mathcal P_+ \theta =0~, \qquad \qquad \mathcal P_\pm = \tfrac{1}{2} (1\pm \Gamma^\star)~ \ , \la{28}\\
&
 \mathcal P_- \theta = \theta~,   \qquad\qquad   \mathcal P_\pm^2 = 1~, \qquad \mathcal P_- \mathcal P_+ = \mathcal P_+ \mathcal P_- =0~.\la{29}
\end{align}
 Then  the expressions in~\rf{24}--\rf{27} take the form (we rescale $\theta \rightarrow \frac{1}{\sqrt{2}} \theta,  \ \bar{\theta} \to \frac{1}{\sqrt{2}} \bar{\theta}$)
\begin{align}
\mathcal L_2 =& - \tfrac{1}{2} \partial_\mu X^I \partial^\mu X^I + i \bar{\theta} \mathcal P_- \Gamma^\mu \partial_\mu \theta~, \qquad \qquad 
\mathcal L_3 =0~, \la{30}\\
\mathcal L_4 =&  -\tfrac{1}{8} \partial_\mu X^I \partial^\mu X^I \partial_\nu X^J \partial^\nu X^J +\tfrac{1}{4} \partial_\mu X^I \partial_\nu X^I \partial^\nu X^J \partial^\mu X^J \no \\
& + \tfrac{i}{4} \partial_\mu X^I \partial^\mu X^I \bar{\theta} \mathcal P_-\Gamma^\nu \partial_\nu \theta - \tfrac{i}{2} \partial_\mu X^I \partial^\nu X^I \bar{\theta} \mathcal P_- \Gamma^\mu  \partial_\nu\theta\no  \\
& - \tfrac{i}{4} \partial_\nu X^I \partial_\lambda X^J \epsilon^{\mu \nu \lambda} \bar{\theta} \mathcal P_-  \Gamma^{IJ} \partial_\mu  \theta + \OO(\theta^4)~.\la{31}
\end{align}
Note that the cubic vertex in~\rf{25}  vanishes after fixing the gauge~\rf{28}.

The  gauge-fixed action describes $\hD=D-3$ free massless bosons  and $\nf $ massless Majorana 3d  fermions, 
with their  numbers of on-shell d.o.f.~being equal for $D=4,5,7,11$, i.e.  $\ha \nf= \hD$  (see Appendix~\ref{A}).
In particular, in the maximal dimension $D=11$   we have  $\hD=8$  bosons and  $\ha \nf=8$    real fermions. 
Their  propagators  are 
\begin{equation}\la{32}
G_B[p] = \frac{1}{p^2-i \varepsilon}~, \qquad \quad G_F[p] = \frac{\mathcal P_- \Gamma^\mu p_\mu}{p^2 - i \varepsilon}~, \qquad \quad \varepsilon >0~.
\end{equation}
 From $\mathcal L_4$  in~\rf{31}  we get the following vertices with  4 bosons,  and with 2 bosons   and   2 fermions
 that we will need below. 
 Labelling the  bosons by the $SO(\hD)$ indices $I_1, I_2, I_3, I_4$ and 
  3-momenta $p_1, p_2, p_3, p_4$ (here with the convention  that  all momenta are  incoming), 
  the  bosonic 4-vertex reads
\begin{align}
\mathcal V_{I_1,I_2,I_3,I_4}[p_1,p_2,p_3,p_4] &=   \Big[ (p_1 \cdot p_2) (p_3 \cdot p_4) -(p_1 \cdot p_3) (p_2 \cdot p_4) -(p_1 \cdot p_4)(p_2 \cdot p_3)  \Big] \delta^{I_1 I_2} \delta^{I_3 I_4}\qquad \qquad  \no \\
&+  \Big[(p_1 \cdot p_3) (p_2 \cdot p_4) -(p_1 \cdot p_2) (p_3 \cdot p_4) -(p_1 \cdot p_4)(p_2 \cdot p_3)\Big] \delta^{I_1 I_3} \delta^{I_2 I_4}\no  \\
&+  \Big[ (p_1 \cdot p_4) (p_2 \cdot p_3) -(p_1 \cdot p_3) (p_2 \cdot p_4) -(p_1 \cdot p_2)(p_3 \cdot p_4) \Big] \delta^{I_1 I_4} \delta^{I_2 I_3}~. \label{33}
\end{align}
For the  vertex  with  two fermions $\bar{\theta}^{\alpha_1}$ and $\theta_{\alpha_2}$ 
with  momenta $p_1, p_2$, and two bosons $X^{I_3}$ and $X^{I_4}$ with  momenta $p_3, p_4$  we get~\foot{The  factor of the projector $\mathcal P_-$ ensures that we are  considering only  the physical components of the  fermions (cf.~\rf{28},~\rf{29}).} 
\begin{equation} \label{34}
\begin{aligned}
 \big(\mathcal V_{I_3,I_4}[p_1,p_2,p_3,p_4]\big)^{\alpha_1}_{\alpha_2} = &\tfrac{1}{2} \delta^{I_3 I_4}  \big(\mathcal P_- \Gamma^\mu\big)^{\alpha_1}_{\alpha_2}  \Big[ p_{2,\mu} (p_3 \cdot p_4) -  p_{4,\mu} (p_2 \cdot p_3) -  p_{3,\mu}(p_2 \cdot p_4)
  \Big]\\
 &+ \tfrac{1}{2} \left(\mathcal P_-\Gamma^{I_3 I_4} \right)^{\alpha_1}_{ \alpha_2} \, \epsilon^{\mu \nu \lambda}  p_{2,\mu} p_{3,\nu} p_{4,\l}~,
\end{aligned}
\end{equation}
with the convention that particles have incoming  momenta, while anti-particles have outcoming momenta
(see Fig.~\ref{f1}).
\begin{figure} \centering
\begin{tikzpicture}[scale=0.9]
\begin{scope}[decoration={
    markings,
    mark=at position 0.5 with {\arrow{latex}}}
    ] 
\draw[postaction={decorate}] (0,0) -- (-1,1) node[left] {$(p_1,\alpha_1)$};
\draw[postaction={decorate}] (-1,-1) node[left] {$(p_2,\alpha_2)$}--(0,0);
\draw[postaction={decorate}] (1,1) node[right]{$(p_3,I_3)$} -- (0,0);
\draw[postaction={decorate}] (1,-1) node[right] {$(p_4,I_4)$} -- (0,0);
\end{scope}
\fill (0,0) circle (0.1); 
\end{tikzpicture} 
\caption{2-fermion  --  2-boson vertex.}
\label{f1}
\end{figure}

From~\rf{33}   we get  the  explicit form 
of the tree-level  amplitude  for  two incoming bosons with labels
  $I_1$, $I_2$ and  $p_1$, $p_2$, and two outgoing bosons with $I_3$,$I_4$ and  $p_3$,$p_4$ 
   (the kinematic variables $s,t,u$ are defined in Appendix~\ref{A}):~\foot{Here 
  we assume that $\hD>1$, i.e.  $D>4$. In the special case of  $D=4$ there is only one transverse boson and the scattering amplitude~\rf{1}  is  given simply  by  $\mathcal M_{D=4} = A + B+C$  so that, in particular, 
  $\mathcal M^{(0)}_{D=4} = -(st + tu + us)$. 
  Note   that  we  will always  do not explicitly indicate 
  the  momentum conservation delta-function factor $\delta^{(3)} (p_1 + p_2 - p_3 -p_4)$. }
 \begin{equation}\la{333}
\mathcal M^{(0)} =- \tfrac{1}{2} \left( t u \, \delta^{I_1 I_2} \delta^{I_3 I_4}  
+  s u \, \delta^{I_1 I_3} \delta^{I_2 I_4} + s t \, \delta^{I_1 I_4} \delta^{I_2 I_3} \right)~.
\end{equation}
Note that   if we restrict   all the 4 momenta to lie in one 2-plane
 (which, at the tree level,  effectively corresponds to the reduction to 2d string theory)
then one can  choose  $s>0,\, t=0,\, u=-s$  and  thus  tree-level scattering becomes diagonal, with only the  transmission coefficient  
$B^{(0)} = \frac{1}{2} s^2$   being non-zero.

\section{One-loop amplitude}
\label{s3}

Let us now  compute the 1-loop scattering amplitude $\mathcal M^{(1)}$ for  four external bosons. 
There are   3    standard   bubble diagrams (Fig.~\ref{f2})  with  either  bosons  or fermions 
propagating in the loop. They are found using the 4-vertices in~\rf{33} and~\rf{34}. 
There are  also tadpole  diagrams  (Fig.~\ref{f3}) 
originating from  the sextic terms $(\del X)^6 +( \del X)^4 (\bar \theta \del \theta)$  in 
$\mathcal L_6$  in~\rf{23} that are discussed in Appendix~\ref{C}.  In the case of  non-compact 
$\mathbb R^{1,2}$  membrane case they  give   only trivial  (cubically divergent) 
contributions  that    may be cancelled by  the local path integral measure contribution 
or automatically  vanish in dimensional regularization that we will  assume.

\begin{figure}
\begin{tikzpicture}[scale=0.65]
\begin{scope}[decoration={
    markings,
    mark=at position 0.5 with {\arrow{latex}}}
    ] 
\draw[postaction={decorate}] (-2,2) node[left] {$(p_1,I_1)$} -- (-1,0);
\draw[postaction={decorate}] (-1,0) to [out=60,in=120] node[midway, above] {$p$}  (1,0);
\draw[postaction={decorate}](1,0) -- (2,2) node[right]{$(p_3,I_3)$};
\draw[postaction={decorate}] (-2,-2) node[left] {$(p_2,I_2)$}--(-1,0);
\draw[postaction={decorate}] (1,0) to [out=-120,in=-60]  node[midway, below] {$q$} (-1,0);
\draw[postaction={decorate}] (1,0) -- (2,-2) node[right] {$(p_4,I_4)$};
\end{scope}
\fill (-1,0) circle (0.1); 
\fill (1,0) circle (0.1); 
\end{tikzpicture} \qquad
\begin{tikzpicture}[scale=0.65]
\begin{scope}[decoration={
    markings,
    mark=at position 0.5 with {\arrow{latex}}}
    ] 
\draw[postaction={decorate}] (-1,2) node[left] {$(p_1,I_1)$}--(0,1);
\draw[postaction={decorate}] (0,1) -- (1,2) node[right] {$(p_3,I_3)$};
\draw[postaction={decorate}] (0,-1) to [out=150,in=-150]  node[midway, left] {$q$} (0,1);
\draw[postaction={decorate}] (0,1) to [out=-30,in=30] node[midway, right] {$p$}  (0,-1);
\draw[postaction={decorate}] (0,-1) -- (1,-2) node[right]{$(p_4,I_4)$};
\draw[postaction={decorate}] (-1,-2) node[left] {$(p_2,I_2)$}--(0,-1);
\end{scope}
\fill (0,1) circle (0.1); 
\fill (0,-1) circle (0.1); 
\end{tikzpicture} \qquad
\begin{tikzpicture}[scale=0.65]

\draw[-latex] (0,1) .. controls (2,2) .. (2.5,-2) node[right] {$(p_4,I_4)$};
\draw[-latex]  (0,-1) .. controls (2,-2) ..  (2.5,2) node[right]{$(p_3,I_3)$};
\begin{scope}[decoration={
    markings,
    mark=at position 0.5 with {\arrow{latex}}}
    ] 
\draw[postaction={decorate}] (-1,2) node[left] {$(p_1,I_1)$}--(0,1);
\draw[postaction={decorate}] (0,-1) to [out=150,in=-150]  node[midway, left] {$q$} (0,1);
\draw[postaction={decorate}] (0,1) to [out=-30,in=30] node[midway, right] {$p$}  (0,-1);
\draw[postaction={decorate}] (-1,-2) node[left] {$(p_2,I_2)$}--(0,-1);
\end{scope}
\fill (0,1) circle (0.1); 
\fill (0,-1) circle (0.1);
\end{tikzpicture} 
\caption{The $s$-, $t$- and $u$-channel one-loop bubble   diagrams}
\label{f2}
\end{figure} 
\begin{figure} \centering
\begin{tikzpicture}[scale=1.3]
\draw[-] (-1,-1) node[left] {$(p_2,I_2)$}--(1,1) node[right] {$(p_3,I_3)$};
\draw[-] (-1,1) node[left] {$(p_1,I_1)$}--(1,-1) node[right] {$(p_4,I_4)$};
\fill (0,0) circle (0.1); 
\draw[->] (0,0) to [out=60,in=0] (0,1) to[out=180,in=120]  (0,0);
\end{tikzpicture} 
\caption{Tadpole  diagram}
\label{f3}
\end{figure}
\subsection{Bosonic loop}

Let us  start with   the contribution of  the  1-loop  diagrams in Fig.~\ref{f2} with only bosons propagating in the loop. 
It is the same as  in the bosonic membrane case  already  considered 
 in~\ci{Seibold:2023zkz}, but we  will repeat  its  computation  here for completeness. 
 The 1-loop  integral corresponding to the $s$-channel   diagram in Fig.~\ref{f2} 
   contains  two bosonic 4-vertices $\mathcal V$
 in~\rf{33}~\foot{The overall factor of $\ha$ comes from the symmetry of the diagram. The two bosons in the loop have   indices $I_p$ and $I_q$, and momenta $p$ and $q$. The conservation of momentum  implies that  $q = p-p_1-p_2 = p-p_3-p_4$.
As in~\rf{333} we  will often  suppress $I_i$ indices in the l.h.s.~of the expressions  below.} 
\begin{equation}
\mathcal M_{B,s}^{(1)} = \frac{1}{2i} \int \frac{d^d p}{(2 \pi)^d}\,  \mathcal V_{I_p,I_q,I_1,I_2}[-p,q,p_1,p_2]\ \mathcal V_{I_p, I_q, I_3, I_4}[p,-q,-p_3,-p_4]\ \frac{1}{p^2 - i \varepsilon}\,  \frac{1}{q^2 - i \varepsilon}~.\la{35}
\end{equation}
Here we assumed  dimensional regularization  but as this integral is UV finite we may set $d=3$. 
To compute this integral we use  the Feynman parametrization 
 \begin{align}
&\mathcal M_{B,s}^{(1)} = \frac{1}{2i} \int_0^1 dx \int \frac{d^d l}{(2 \pi)^d}\  \frac{1}{(l^2+\Delta_s)^2}  \ N_{B,s}^{(1)}~,
\la{36}\\  &
l = p - x(p_1+p_2)~, \qquad\qquad   \Delta_s \equiv  - x(1-x) s - i \varepsilon ~,\la{37}
\\ &
N_{B,s}^{(1)} = \mathcal V_{I_p,I_q,I_1,I_2}[-p,q,p_1,p_2]\ \mathcal V_{I_p, I_q, I_3, I_4}[p,-q,-p_3,-p_4]~.\la{38}
\end{align}
Expanding $N_{B,s}^{(1)}$ in powers of $l$ we  observe that odd  powers  can be dropped as they do not 
 contribute to the integral    so that
 \begin{align}
N_{B,s}^{(1)} &= N_0 + N_2 l^2 + N_{\mu \nu} l^\mu l^\nu  + N_4 (l^2)^2 + N'_{\mu \nu} l^2 l^\mu l^\nu + N_{\mu \nu \rho \sigma} l^\mu l^\nu l^\rho l^\sigma~,\la{39}
\\ 
N_{\mu \nu \rho \sigma} &= 4 \hD \delta^{I_1 I_2} \delta^{I_3 I_4} p_{1,\mu} p_{2,\nu} p_{3,\rho} p_{4,\sigma}~,\qquad 
N'_{\mu \nu} = (\hD-2)\delta^{I_1 I_2} \delta^{I_3 I_4} s \left( p_{1,\mu} p_{2,\nu} + p_{3,\mu} p_{4,\nu}  \right)~,\no \\
N_4 &=\tfrac{1}{4} \left(( \hD-4) \delta^{I_1 I_2} \delta^{I_3 I_4} + 2 \delta^{I_1 I_3} \delta^{I_2 I_4}+ 2 \delta^{I_1 I_4} \delta^{I_2 I_3} \right) {s^2} ~,\no\\
N_2 &= -x(1-x) \left( \delta^{I_1 I_2} \delta^{I_3 I_4} - \delta^{I_1 I_3} \delta^{I_2 I_4}  - \delta^{I_1 I_4} \delta^{I_2 I_3}   \right)s^3 ~,\no \\
N_{\mu \nu} &= -2 x(1-x) \delta^{I_1 I_2} \delta^{I_3 I_4} s^2 (p_{1,\mu} p_{2,\nu} + p_{3,\mu} p_{4,\nu} )\no \\
&\qquad + \delta^{I_1 I_3} \delta^{I_2 I_4} s^2 \big[p_{1,\mu} p_{3,\nu} + p_{2,\mu} p_{4,\nu}-2 x(1-x) (p_{1,\mu} + p_{2,\mu})(p_{3,\nu} + p_{4,\nu}) \big] \no \\
&\qquad + \delta^{I_1 I_4} \delta^{I_2 I_3} s^2 \big[p_{1,\mu} p_{4,\nu} + p_{2,\mu} p_{3,\nu}-2 x(1-x) (p_{1,\mu} + p_{2,\mu})(p_{3,\nu} + p_{4,\nu}) \big]~,\no  \\
N_0 &= \half x^2 (1-x)^2  \left( \delta^{I_1 I_3}\delta^{I_2 I_4} + \delta^{I_1 I_4} \delta^{I_2 I_3} \right) {s^4}~.\la{40}
\end{align}
Computing the integrals over the momentum $l$ and the Feynman parameter $x$ then  leads to
\begin{equation}\la{41}
\begin{aligned}
\mathcal M_{B,s}^{(1)} &= \tfrac{1}{256} (-s)^{3/2} \Big[\delta^{I_1 I_2} \delta^{I_3 I_4} \Big( \big(\tfrac{3 \hD}{32}-1\big) s^2 - \tfrac{\hD}{4}  t u  \Big) \\
&\hspace{2.6 cm}+ \delta^{I_1 I_3} \delta^{I_2 I_4} s(3s+2t) + \delta^{I_1 I_4} \delta^{I_2 I_3} s(3s+2 u) \Big]~.
\end{aligned}
\end{equation}
The  amplitudes in the $t$- and $u$-channels are obtained through crossing  (cf.~\rf{11}) 
\begin{equation}
\mathcal M_{B,t}^{(1)} = \left. \mathcal M_{B,s}^{(1)} \right|_{s \leftrightarrow t, I_2 \leftrightarrow I_3}~, \qquad \mathcal M_{B,u}^{(1)} = \left. \mathcal M_{B,s}^{(1)} \right|_{s \leftrightarrow u, I_2 \leftrightarrow I_4}~. \la{42}
\end{equation}
The  final expression for the  bosonic 1-loop amplitude  is given by 
\begin{align}
\mathcal M_B^{(1)}  =& \mathcal M_{B,s}^{(1)} +  \mathcal M_{B,t}^{(1)} +  \mathcal M_{B,u}^{(1)} \la{55}\\
=&\tfrac{1}{256}  \delta^{I_1 I_2} \delta^{I_3 I_4}   \Big[(-s)^{3/2}   \left(\big(\tfrac{3\hD}{32}-1\big) s^2 - \tfrac{\hD}{4} t u  \right) + (-t)^{3/2}  t(3t+2s)  + (-u)^{3/2} u(3u+2 s) \Big]\no\\
 &+ \tfrac{1}{256} \delta^{I_1 I_3} \delta^{I_2 I_4} \Big[ (-t)^{3/2}  \left( \big(\tfrac{3 \hD}{32}-1\big) t^2 - \tfrac{\hD}{4} s u  \right) +  (-s)^{3/2} s(3s+2t) + (-u)^{3/2} u(3u+2t) \Big] \no\\
 &+\tfrac{1}{256} \delta^{I_1 I_4} \delta^{I_2 I_3}  \Big[ (-u)^{3/2} \left( \big(\tfrac{3 \hD}{32}-1\big) u^2 - \tfrac{\hD}{4} s t  \right) + (-s)^{3/2} s(3s+2 u) + (-t)^{3/2} t(3t+2 u)  \Big]~.\no
\end{align}
Note that when restricted  to planar scattering  with  $t=0, \ u=-s$ this gives 
\begin{align}
\mathcal M_B^{(1)}\Big|_{t=0} =&  \tfrac{1}{256} \delta^{I_1 I_2} \delta^{I_3 I_4}   \Big[(-s)^{3/2}   \big(\tfrac{3}{32}\hD-1\big)  + s^{3/2} \Big] s^2 \no  \\
 & + \tfrac{3}{256} \delta^{I_1 I_3} \delta^{I_2 I_4} \Big[ (-s)^{3/2}  + s^{3/2} \Big] s^2 
  +\tfrac{1}{256} \delta^{I_1 I_4} \delta^{I_2 I_3}  \Big[ s^{3/2} \big(\tfrac{3}{32}\hD-1\big) + (-s)^{3/2}   \Big]
   s^2~.\la{43}
\end{align}
 Note that compared to the  case of the  bosonic string (which  corresponds to the  double dimensional reduction of the membrane), where the 1-loop scattering  amplitude took special form   for  $D=26$~\ci{Dubovsky:2012wk}
 here there is no  special value of $D=\hD+3$ for which this 
 expression simplifies.

\subsection{Fermionic  loop}

The $s$-channel  diagram in Fig.~\ref{f2}   with fermions   propagating in the loop  gives  the following Feynman integral
\begin{align}
\mathcal M_{F,s}^{(1)} 
&= i\sum_{\alpha_1, \alpha_2, \alpha_3, \alpha_4} \int \frac{d^dp}{(2 \pi)^d} \big(\mathcal V_{I_1,I_2}[p, q,p_1,p_2]\big)^{\alpha_1}_{\alpha_2} \ \big(\mathcal V_{I_3, I_4}[q,p,-p_3, -p_4]\big)^{\alpha_4}_{\alpha_3}\  \frac{(\mathcal P_-\Gamma^\mu)^{\alpha_1}_{\alpha_3}  p_\mu \, (\mathcal P_-\Gamma^\nu)^{\alpha_4}_{\alpha_2}  q_\nu}{(p^2 - i \varepsilon) (q^2- i \varepsilon)}\no \\
&= i \int \frac{d^dp}{(2 \pi)^d} \text{Tr} \Big( \mathcal V_{I_1 I_2}[p,q,p_1,p_2]\ \frac{\mathcal P_-\Gamma^\mu q_\mu}{ q^2-i \varepsilon}\  \mathcal V_{I_3 I_4}[q,p,-p_3,-p_4] \ \frac{ \mathcal P_-\Gamma^\nu p_\nu}{ p^2-i \varepsilon} \Big)~.
\la{44}\ \end{align}
Here  $\alpha_1, \alpha_2, \alpha_3, \alpha_4$ are the spinor indices of the fermions in the 4-vertices in~\rf{34} and the propagators in~\rf{32}. The  sum over them  
 produces  a trace of  products of $\Gamma$-matrices. The integral is again UV finite so we may set $d=3$. 
As in~\rf{36}--\rf{38}   we also get 
\begin{align}
\mathcal M_{F,s}^{(1)} &= {i}\int_0^1 dx \int \frac{d^dl}{(2 \pi)^d}  \frac{1}{(l^2+\Delta_s)^2}  N_{F,s}^{(1)}~, 
\qquad l = p - x(p_1+p_2)~, \qquad \Delta_s = - x(1-x) s - i \varepsilon ~,\no \\
N_{F,s}^{(1)} &= \text{Tr}\Big(   \mathcal V_{I_1 I_2}[p,q,p_1,p_2]\,  \mathcal P_-\Gamma^\mu q_\mu\,  \mathcal V_{I_3 I_4}[q,p,-p_3,-p_4]\,  \mathcal P_-\Gamma^\nu p_\nu \Big)~.\la{45}
\end{align}
The analog of~\rf{39}  here is 
  (we use the relations for $\Gamma$-matrix traces   from  Appendix~\ref{A})
\begin{align}
N_{F,s}^{(1)}&= N_0 + N_2 l^2 + N_{\mu \nu} l^\mu l^\nu + N_4 (l^2)^2 + N'_{\mu \nu} l^2 l^\mu l^\nu + N_{\mu \nu \rho \sigma} l^\mu l^\nu l^\rho l^\sigma~,\la{46}
\\ 
  N_{\mu \nu \rho \sigma} & = \nf  \delta^{I_1 I_2} \delta^{I_3 I_4} p_{1,\mu} p_{2,\nu} p_{3,\rho} p_{4,\sigma}~,\qquad \ \ \ 
N_4 = \tfrac{1}{32}\nf  \delta^{I_1 I_2} \delta^{I_3 I_4} s^2 ~, \no \\
N'_{\mu \nu} &= \tfrac{1}{16}\nf \Big(  \delta^{I_1 I_2} \delta^{I_3 I_4}  \Big[ 2s (   p_{1,\mu} p_{2,\nu} + p_{3,\mu} p_{4,\nu} )- t (p_{1,\mu} p_{3,\nu}+p_{2,\mu} p_{4,\nu} ) - u ( p_{1,\mu} p_{4,\nu}+p_{2,\mu} p_{3,\nu}) \Big]\no  \\
&\qquad - 2 \left( \delta^{I_1 I_3} \delta^{I_2 I_4} - \delta^{I_1 I_4} \delta^{I_2 I_3} \right) \epsilon^{\mu \lambda_1 \lambda_2} \epsilon^{\nu \lambda_3 \lambda_4} p_{1,\lambda_1} p_{2,\lambda_2} p_{3,\lambda_3} p_{4,\lambda_4}\Big)~, \no \\
N_{\mu \nu} &= \tfrac{1}{32}\nf \Big\{  x(1-x) \delta^{I_1 I_2} \delta^{I_3 I_4} s^2 \big(  p_{1,\mu} p_{1,\nu} + p_{2,\mu} p_{2,\nu} +  p_{3,\mu} p_{3,\nu} +  p_{4,\mu} p_{4,\nu}  \big) \no\\
& \ +2 x(1-x)\delta^{I_1 I_2} \delta^{I_3 I_4} s \big( t p_{2,\mu} p_{3,\nu} + u p_{2,\mu} p_{4,\nu} + u p_{1,\mu} p_{3,\nu} + t p_{1,\mu} p_{4,\nu} -s p_{1,\mu} p_{2,\nu} - s p_{3,\mu} p_{4,\nu} \big)\no \\
&\ - 4   x (1-x) \left( \delta^{I_1 I_3} \delta^{I_2 I_4} - \delta^{I_1 I_4} \delta^{I_2 I_3} \right)  s \epsilon^{\mu \lambda_1 \lambda_2}\epsilon^{\nu \lambda_3 \lambda_4} p_{1,\lambda_1} p_{2,\lambda_2} p_{3,\lambda_3} p_{4,\lambda_4} \Big\}~, \no  \\
N_2 &= -\tfrac{1}{32}\nf \,  x(1-x) \delta^{I_1 I_2} \delta^{I_3 I_4}  s^3 ~, \qquad N_0 =0~.\la{47}
\end{align}
Here  $\nf $ is the number of  real spinor components  after $\k$-symmetry gauge fixing
(see~\rf{a7},\rf{a8}). 
To account for  the Majorana  condition on $\theta$  we assume that in~\rf{45}  $\Tr $   includes  an extra $\half$ factor, so  that  effectively  in~\rf{44}   we have $\Tr[ \mathcal P_-]\to \ha  \nf $ (see also  Appendix~\ref{A}). 

The integrals over the loop momentum $l$ and the Feynman parameter $x$ then give (cf.~\rf{41})
\begin{equation}
\mathcal M_{F,s}^{(1)} = \tfrac{1}{256} \tfrac{1}{64} \nf\, (-s)^{3/2}\Big[ \delta^{I_1 I_2} \delta^{I_3 I_4} \left( s^2 -8 t u \right)  + \delta^{I_1 I_3} \delta^{I_2 I_4} 4  s (s+2t) + \delta^{I_1 I_4} \delta^{I_2 I_3} 4  s(s+2 u) \Big]. \la{48}
\end{equation}
The  $t$- and $u$-channel  expressions  are found   using   crossing  relations as in~\rf{42}. 
The  total contribution from the fermionic loop is then 
\begin{align}
\mathcal M_F^{(1)}&  = \mathcal M_{F,s}^{(1)} + \mathcal M_{F,t}^{(1)} + \mathcal M_{F,u}^{(1)}\la{49}  \\
&=\tfrac{1}{256} \tfrac{1}{64}\nf  \Big\{  \delta^{I_1 I_2} \delta^{I_3 I_4}   \left[(-s)^{3/2}   \left(s^2 - 8 t u  \right) + (-t)^{3/2} 4    t(t+2s)  + (-u)^{3/2} 4  u(u+2 s) \right]\no  \\
 &\qquad\qquad \  +  \delta^{I_1 I_3} \delta^{I_2 I_4} \left[ (-t)^{3/2}  \left[  t^2 - 8 s u  \right) +  (-s)^{3/2} 4  s(s+2t) + (-u)^{3/2} 4 u(u+2t) \right]\no  \\
 &\qquad\qquad \ + i\,  \delta^{I_1 I_4} \delta^{I_2 I_3}  \left[ (-u)^{3/2} \left(  u^2 - 8 s t  \right) + (-s)^{3/2} 4 s(s+2 u) + (-t)^{3/2} 4  t(t+2 u)  \right]\Big\}.\no 
\end{align}

\subsection{Total one-loop amplitude}

Adding together  the bosonic loop~\rf{43} and  the fermionic loop~\rf{49}  contributions we get  
\begin{align}
\mathcal M^{{(1)}} = \mathcal M_B^{(1)} + \mathcal M_F^{(1)}  = &\tfrac{1}{256} \delta^{I_1 I_2} \delta^{I_3 I_4} 
  \Big\{(-s)^{3/2} \Big[ \big( \tfrac{3\hD}{32}-1+ \tfrac{\nf }{64} \big) s^2 - \big( \tfrac{\hD}{4}+  \tfrac{\nf }{8} \big) t u \Big] \no  \\
&+ (-t)^{3/2}  \Big[ ( 3 +\tfrac{\nf }{16})   t^2 + (2 + \tfrac{\nf }{8})  st \Big] 
+ (-u)^{3/2} \Big[ ( 3 +\tfrac{\nf }{16})   u^2+ (2 + \tfrac{\nf }{8})  su \Big]  \Big\} \no \\
&\qquad + (s \leftrightarrow t, I_2 \leftrightarrow I_3) + (s \leftrightarrow u, I_2 \leftrightarrow I_4)~.\la{50}
\end{align}
Here $\nf = 2 \hD=2 (D-3) $  for the  classically  allowed   $D=7,5,4, 11$ supermembrane cases.
We observe  that a drastic simplification happens   in   the maximal allowed dimension $D=11$ where 
the 1-loop amplitude~\rf{50} becomes proportional 
to the tree-level one~\rf{333}, i.e.   
\begin{equation}  \label{51}
\begin{aligned}
\mathcal M^{(1)} &= - \tfrac{1}{64}  \Big[ (-s)^{3/2} + (-t)^{3/2} + (-u)^{3/2} \Big] \Big( \delta^{I_1 I_2} \delta^{I_3 I_4} tu  + \delta^{I_1 I_3} \delta^{I_2 I_4} su   + \delta^{I_1 I_4} \delta^{I_2 I_3} st  \Big) \\
&= 
 \tfrac{1}{32}  \Big[ (-s)^{3/2} + (-t)^{3/2} + (-u)^{3/2} \Big]  \mathcal M^{(0)}~.
\end{aligned}
\end{equation}
This is the expression~\rf{6}    discussed in the Introduction
with the total amplitude given in~\rf{7}.

\section{Massless boson scattering on  $\mathbb R^{1,1}\times S^1$ supermembrane}
\label{s4}

Let us now  consider  scattering of massless bosonic 2d modes 
in the $\mathbb R^{1,1}\times S^1$ supermembrane   vacuum generalizing the bosonic membrane  computation in~\ci{Seibold:2023zkz}.  
   Using the corresponding static gauge  and Fourier   expanding 
in the $S^1$ coordinate  we get an effective 2d model  containing  massless fields of the  superstring  sector 
 plus a tower of massive  2d  modes. Our aim will be to compute the 1-loop   scattering  amplitude  for the 2d massless  bosons
 and to show its simplification in the special case of the $D=11$ supermembrane.

\subsection{Effective two-dimensional action}

We  will assume that one of the  target-space directions is a circle, e.g.,  
 $X^2 \in (-\pi\rR,+\pi \rR]$ 
 and that  the membrane  wraps this $S^1$ direction. 
 The corresponding static gauge  is~\rf{21} 
 where $\s^2$ is now compact. 
Expanding the  transverse fields  $X^I$  in Fourier modes in $\s^2$  we get 
\begin{equation}\la{511}
X^I(\sigma^0, \sigma^1, \sigma^2) = \sum_{n=-\infty}^{\infty} X_n^I(\sigma^0,\sigma^1) \,e^{\frac{ i n}{\rR} \sigma^2}~, \qquad I=3,\dots,D-1~, \qquad X_{-n}^I = (X_n^I)^*~.
\end{equation}
For the fermions   which are  world-volume scalars we  get,    assuming  (as required by global supersymmetry) 
periodicity in $\s^2$ 
\begin{equation}\la{52}
\theta (\sigma^0, \sigma^1, \sigma^2) = \sum_{n=-\infty}^{\infty} \theta_n(\sigma^0,\sigma^1) \,e^{\frac{ i n}{\rR} \sigma^2}~, \qquad \qquad \bar{\theta} (\sigma^0, \sigma^1, \sigma^2) = \sum_{n=-\infty}^{\infty} \bar{\theta}_{-n}(\sigma^0,\sigma^1) \,e^{\frac{ i n}{\rR} \sigma^2}~.
\end{equation}
The fermion field $\theta$ and thus each  mode $\theta_n$  is  again   subject to the 
  $\k$-symmetry gauge condition~\rf{28}. 
 Starting with the   supermembrane action in the static gauge~\rf{13}, integrating over 
  $\sigma^2$ and expanding in powers of the fields as in~\rf{23}   we  arrive at the following  action 
  for a tower of massless and massive 2d fields (here $d^2\s = d\s^0 d\s^1$  and $\hat \m =0,1$)~\foot{In more familiar notation 
   in the case of  $D=11$ the labelling of the  compact  coordinate is 
  $X_2 \to X_{10}$ and  thus $\Gamma_2 \to \Gamma_{10}$.}
   \begin{align}\la{53}
&S = 2 \pi \rR\,  T_2  \int d^2 \sigma \, \hat{\mathcal L}~, \qquad\qquad \ \  \hat{\mathcal L} = -1 + \hat{\mathcal L}_2 + \hat{\mathcal L}_4 + \dots~, \\
&\hat{\mathcal L}_2 = \sum_{n=- \infty}^{\infty} \Big( -\tfrac{1}{2} \partial_{\hat{\mu}} X_{-n}^I \partial^{\hat{\mu}} X_n^I - \tfrac{1}{2} m^2_n  X_{-n}^I X_n^I + i \bar{\theta}_n \mathcal P_- \Gamma^{\hat{\mu}}  \partial_{\hat{\mu}} \theta_n - m_n \bar{\theta}_n \mathcal P_-\Gamma_2  \theta_n  \Big)~,  \la{54}\\ 
&\qquad \qquad \qquad \qquad  m_n = {n\ov \rR} \ . \la{544}
\end{align}
 The free  equations of motion are
$
\big( \partial_{\hat{\mu}} \partial^{\hat{\mu}} - m^2_n  \big) X_n^I =0$ and $ \big( i \Gamma^{\hat{\mu}} \partial_{\hat{\mu}} - m_n \Gamma_2 \big) \mathcal P_- \theta_n =0$.
Squaring  the fermionic equation of motion one gets $\big( \partial_{\hat{\mu}} \partial^{\hat{\mu}} - m^2_n \big) \mathcal P_- \theta_n =0$ consistent with an effective 2d supersymmetry. 
Note that  $\hat{\Gamma}=\mathcal P_-\Gamma_2 \mathcal P_-= \mathcal P_-\Gamma_2 $ satisfies $(\hat{\Gamma})^2 = \mathcal P_-$.  
The propagators are  given by (cf.~\rf{32}) 
\begin{equation}\la{555}
\hat{G}_B[p] = \frac{1}{p^2 + m_n^2 - i \varepsilon}~, \qquad\qquad  \hat{G}_F[p] = \mathcal P_-\frac{ \Gamma^{\hat{\mu}}  \,p_{\hat{\mu}} + m_n\Gamma_2}{p^2 + m_n^2 - i \varepsilon} ~.
\end{equation}
For the quartic interaction term in~\rf{53}  we  get 
\begin{align}
& \hat{\mathcal L}_4 = \hat{\mathcal L}_4^{(B)} + \hat{\mathcal L}_4^{(F)} \ ,\la{56} \\
\mathcal L_4^{(B)} =\tfrac{1}{8}  \sum_{n_1, n_2, n_3, n_4} \Big( &-
 \partial_{\hat{\mu}} X_{n_1}^I \partial^{\hat{\mu}} X_{n_2}^I \partial_{\hat{\nu}} X_{n_3}^J \partial^{\hat{\nu}} X_{n_4}^J +  2\partial_{\hat{\mu}} X_{n_1}^I \partial_{\hat{\nu}} X_{n_2}^I \partial^{\hat{\nu}} X_{n_3}^J \partial^{\hat{\mu}} X_{n_4}^J \no \\
& + 2 {n_3 n_4} \partial_{\hat{\mu}} X_{n_1}^I \partial^{\hat{\mu}} X_{n_2}^I X_{n_3}^I X_{n_4}^I - 4{n_2 n_3}
\partial_{\hat{\mu}} X_{n_1}^I X_{n_2}^I X_{n_3}^J \partial^{\hat{\mu}} X_{n_4}^J \no \\
&+{n_1 n_2 n_3 n_4} X_{n_1}^I X_{n_2}^I X_{n_3}^J X_{n_4}^J \Big) \delta(n_1+n_2+n_3+n_4)~, \la{57} \\
\mathcal L_4^{(F)} =  \tfrac{i}{4}
\sum_{n_1, n_2, n_3, n_4} \Big( & \partial_{\hat{\mu}} X_{n_1}^I \partial^{\hat{\mu}} X_{n_2}^I \bar{\theta}_{-n_3}
 \mathcal P_- \Gamma^{\hat{\nu}} \partial_{\hat{\nu}} \theta_{n_4} - 2
 \partial_{\hat{\mu}} X_{n_1}^I \partial_{\hat{\nu}} X_{n_2}^I \bar{\theta}_{-n_3} \mathcal P_- \Gamma^{\hat{\nu}}  \partial^{\hat{\mu}} \theta_{n_4}\no \\
&\te  + \frac{i n_4}{\rR} \partial_{\hat{\mu}} X_{n_1}^I \partial^{\hat{\mu}} X_{n_2}^I \bar{\theta}_{-n_3} \mathcal P_-\Gamma_2  \theta_{n_4}
+  2  \frac{ n_2 n_4}{\rR^2} 
\partial_{\hat{\mu}} X_{n_1}^I X_{n_2}^I \bar{\theta}_{-n_3} \mathcal P_- \Gamma^{\hat{\mu}}  \theta_{n_4}\no  \\
&\te -  \frac{i n_4}{\rR} \partial_{\hat{\mu}} X^I_{n_1} \partial_{\hat{\nu}} X^J_{n_2} \varepsilon^{\hat{\mu} \hat{\nu}} \bar{\theta}_{-n_3} \mathcal P_- \Gamma^{IJ}  \theta_{n_4}
 -\frac{ n_1 n_2}{\rR^2} X_{n_1}^I X_{n_2}^I \bar{\theta}_{-n_3} \mathcal P_- \Gamma^{\hat{\nu}}  \partial_{\hat{\nu}} \theta_{n_4}\no \\
&\te - 2 \frac{i n_2}{\rR} \partial_{\hat{\mu}} X_{n_1}^I X_{n_2}^I \bar{\theta}_{-n_3} \mathcal P_-\Gamma_2\partial^{\hat{\mu}} \theta_{n_4}  + 2 \frac{i n_1}{\rR} X_{n_1}^I \partial_{\hat{\nu}} X_{n_2}^J \varepsilon^{ \hat{\mu}  \hat{\nu}} \bar{\theta}_{-n_3} \mathcal P_- \Gamma^{IJ}  \partial_{\hat{\mu}} \theta_{n_4} \no \\
&\te
+  \frac{i n_1 n_2 n_4}{\rR^3} X_{n_1}^I X_{n_2}^I \bar{\theta}_{-n_3} \mathcal P_-\Gamma_2  \theta_{n_4} \Big)
 \,  \delta(n_1+n_2+n_3+n_4)\  +  \ \OO(\theta^4)  .\la{58}
\end{align}
Here $ \varepsilon^{\hat{\mu}  \hat{\nu}} =  \epsilon^{\hat{\mu}  \hat{\nu}2 }$ is the 2d antisymmetric tensor. 
The Feynman rules for the vertices is  obtained from~\rf{33} and~\rf{34} by the replacement (here $j,k=1,2,3,4$
and $v_\mu=(v_0,v_1,v_2)$ is a  3-vector)
\begin{equation}\la{59} 
p_j \cdot p_k \rightarrow p_j \cdot p_k + \frac{n_j n_k}{\rR^2}~, \qquad \qquad  v \cdot p_j \rightarrow v \cdot p_j + v_2  \frac{n_j}{\rR} ~. 
\end{equation}
 The momenta $p_j$ in the r.h.s of these expressions are  the  2d  ones. 
In addition, we will need to  ensure the conservation of the 2d momentum and the mode number at each  vertex. 
 
\subsection{One-loop amplitude}
We will focus on the 4-point  amplitude of   massless 2d  bosons, i.e.~the  bosonic  modes  with 
\begin{equation}
n_1 = n_2 = n_3 = n_4 =0~,\la{60}
\end{equation} 
  with all  massive bosonic and fermionic  modes  propagating   in the loops.
The mode number conservation then implies that the two  massive propagators in the loops in diagrams in Fig. \ref{f2} 
 have equal mode  numbers or masses. 

The  contribution to the  1-loop  amplitude coming from the  bosonic  loop was already  calculated in~\ci{Seibold:2023zkz}. 
As discussed in~\ci{Seibold:2023zkz}, while the  3d  theory on $\mathbb R^{1,2}$  is manifestly free of log UV divergences, 
if one first  compactifies  one dimension and represents the theory as  the   one for an infinite  tower of 2d modes then 
logarithmic divergences   may disappear  only after  one   performs the summation over the $S^1$ modes  with an appropriate 
regularization.\foot{Viewed  as a 2d theory, the  3d theory on $S^1$   must still have  the same  UV behaviour
 as  in the  uncompactified  case. To preserve this equivalence requires an appropriate choice of the regularization that is consistent with restoration of the 3d Lorentz symmetry in the UV limit.}
As in~\ci{Seibold:2023zkz} we will use  a combination of dimensional  regularization near $d=2$   and the Riemann $\zeta$-function for the sum over the mode number $n$. 

One finds that  for fixed  mode  number $n$  field in the  loop there is a 2d  divergent contribution  (using dimensional regularization with $d=2-2\epsilon$)
\begin{align}
\hat{\cal M }_{B,n,\epsilon}^{(1)}& =\te  -\tfrac{1}{96 \pi} \Big(\tfrac{1}{\epsilon} - \gamma + \ln 4 \pi \Big)  
\Big\{ (\hD-6)\,  stu\,  \left( \delta^{I_1 I_2} \delta^{I_3 I_4} + \delta^{I_1 I_3} \delta^{I_2 I_4} + \delta^{I_1 I_4} \delta^{I_2 I_3} \right)\no \\
&\te   \quad + 6n^2\Big[ (2 s^2 - \hD tu) \delta^{I_1 I_2} \delta^{I_3 I_4}  
+  (2 t^2 - \hD su) \delta^{I_1 I_3} \delta^{I_2 I_4} 
+ (2 u^2 - \hD s t ) \delta^{I_1 I_4} \delta^{I_2 I_3} \Big] \Big\}~.\la{61}
\end{align}
Here the $s,t,u$ are defined in terms of 2d  massless  external momenta so we may  choose  $s>0$, $t=0$, $u=-s$
which is assumed below. 
There is  also  a finite contribution  given by~\foot{$Q_n(s) $  originates from  the standard massive  scalar 
1-loop integral  
$\int {d^2q \ov (2\pi)^2} \,  {1\ov q^2 +n^2} \, {1\ov (q-p_1-p_2)^2  + n^2}$.}
\begin{align}
\hat{\cal M }_{B,n}^{(1)} &=  \hat{A}^{(1)}_{B,n}      \delta^{I_1 I_2} \delta^{I_3 I_4} + 
\hat{B}^{(1)}_{B,n} \delta^{I_1 I_3} \delta^{I_2 I_4} + \hat{C}^{(1)}_{B,n} \delta^{I_1 I_4} \delta^{I_2 I_3}     \ ,     \la{62}
\\
\hat{A}^{(1)}_{B,n} &=\te  - \frac{s^2}{192 \pi} \Big[(\hD-24) s  + 
6    (\hD+4) n^2  - 24 n^2  \ln n^2 
- 6  n^2     (\hD n^2 - 2 s) Q_n(-s)  + 12 n^2 s Q_n(s) \Big]~,\no \\
\hat{B}^{(1)}_{B,n} &= \te \frac{s^2}{192 \pi} \Big[ - 12 \hD n^2  + 12 \hat D n^2  \ln n^2 
  +   6 s(s-2 n^2) Q_n(-s) + 6 s (s+2 n^2) Q_n(s)  \Big]~,\la{63} \\
\hat{C}^{(1)}_{B,n} &=\te  \frac{s^2}{192 \pi} \Big[(\hD-24) s - 6 
  (\hD+4) n^2   + 24 n^2  \ln n^2 
  + 6 n^2 (\hD n^2 +2 s)  Q_n(s)  +12 n^2  s Q_n(-s)  \Big]~,\no 
  \\ & \qquad \qquad 
Q_n(s) \equiv \te   - \frac{2}{s \sqrt{1+\frac{4 n^2}{s}}} \ln   \frac{ \sqrt{1+\frac{4 n^2}{s}}-1}{\sqrt{1+\frac{4 n^2}{s}}+1}\ .\la{64}
\end{align}
For simplicity in~\rf{61}--\rf{64}  we set $\rR=1$  but  the dependence on it can be restored by $n \to { n \over \rR}$. 

To find the fermionic loop contribution  we need the 2-boson -- 2-fermion vertex following from~\rf{58} (cf.~\rf{34}).
We
 shall assume that the  spinor  and 2d antisymmetric tensor algebra is 
done in $d=2$ and then 
 the resulting   momentum  integral is continued to $d=2 -2 \epsilon$. We  find  the UV divergent part 
given by (cf.~\rf{61}) 
\begin{equation}\la{65}
\begin{aligned}
\hat {\cal M}_{F,n,\epsilon}^{(1)} &= \te \frac{1}{96 \pi} \tfrac{1}{4}\, \nf  \Big( \frac{1}{\epsilon} - \gamma + \ln 4 \pi \Big) \Big\{-s t u \left( \delta^{I_1 I_2} \delta^{I_3 I_4} + \delta^{I_1 I_3} \delta^{I_2 I_4} + \delta^{I_1 I_4} \delta^{I_2 I_3} \right) \\ 
&\qquad\qquad  + 3 n^2  \Big[s^2-2 tu + 4 (p_1 \wedge p_4\  p_2 \wedge p_3 + p_1 \wedge p_3\  p_2 \wedge p_4) \Big]  \delta^{I_1 I_2} \delta^{I_3 I_4} \\
&\qquad \qquad +  3 n^2 \Big[ t^2 -2 su+4 (p_1 \wedge p_4 \ p_3 \wedge p_2 + p_1 \wedge p_2 \ p_3 \wedge p_4) \Big] \delta^{I_1 I_3} \delta^{I_2 I_4} \\
&\qquad \qquad  + 3 n^2 \Big[ u^2-2 s t  + 4 (p_1 \wedge p_2 \ p_4 \wedge p_3 + p_1 \wedge p_3\  p_4 \wedge p_2) \Big] \delta^{I_1 I_4} \delta^{I_2 I_3} \Big\}~.
\end{aligned}
\end{equation}
Here $p_1 \wedge p_2 =  \ve^{\hat \mu \hat \nu} \, p_{1,\hat \mu}\,  p_{2,\hat \nu}$, etc.  
The divergent terms in~\rf{61} and~\rf{65} will vanish   after the summation over $n$ (see below).

In the expression for the   finite part we may   again  use the 2d   kinematics constraint
 $t=0, u=-s$ which corresponds to choosing $p_1 = p_3$ and $p_2 = p_4$ (with the convention that particles  1,2 
  are  incoming
 and 3,4 are outgoing),  so that, in particular,  that  $p_1 \wedge p_3 =0$  and $p_1 \wedge p_2 = p_1 \wedge p_4$. 
 Then we get  the following  fermionic loop counterparts of  the coefficients~\rf{63}  in~\rf{62}
\begin{equation} \la{66}
\begin{aligned}
\hat{A}_{F,n}^{(1)} &= \te \frac{1}{192 \pi} \tfrac{1}{4}\nf \, s^2 \Big[ - s + 18 n^2 + 3 n^2  (s-4 n^2) Q_n(-s) + 3 n^2 s Q_n(s)   \Big]~, \\
\hat{B}_{F,n}^{(1)} &= \te \frac{1}{192 \pi} \frac{3 }{4}\nf  \, s^2\, n^2 \Big[4 - 8 \ln n^2 + s Q_n(-s) -s Q_n(s)  \Big]~, \\
\hat{C}_{F,n}^{(1)} &=\te  \frac{1}{192 \pi} \tfrac{1}{4}\nf\, s^2 \Big[  s + 18 n^2 - 3 n^2  s Q_n(-s) - 3 n^2 (s + 4 n^2) Q_n(s)   \Big]~.
\end{aligned}
\end{equation}
Adding the finite  parts of the bosonic~\rf{63} and the fermionic~\rf{65}  1-loop amplitudes  gives 
\begin{equation}\la{67}
\begin{aligned}
\hat{A}_{n}^{(1)} &= - \tfrac{1}{192 \pi} s^2\Big[\te  \big(\hD-24 + \tfrac{1}{4}\nf \big)s + 6 \big(\hD + 4 - \frac{3 \nf }{4}\big) n^2 - 24 n^2 \ln n^2 \\
&\qquad\qquad \ \ \te  -6 n^2 \big[(\hD- \frac{\nf }{2}) n^2 - 2 s (1- \frac{\nf }{16}) \big] Q_n(-s)  + 12 n^2 s(1- \frac{\nf }{16}) Q_n(s) \Big]~,\\
\hat{B}_n^{(1)} &=  \tfrac{1}{192 \pi} s^2\Big[\te -12 (\hD - \tfrac{1}{4}\nf) n^2 + 12 (\hD - \frac{\nf }{2}) n^2 \ln n^2 \\
&\qquad\qquad \ \   +6 s\big(s - 2 n^2 + \tfrac{\nf }{8} n^2\big) Q_n(-s) + 6 s\big(s + 2 n^2 - \tfrac{\nf }{8} n^2\big) Q_n(s)\Big]~, \\
\hat{C}_{n}^{(1)} &=  \tfrac{1}{192 \pi} s^2\Big[\te \big(\hD-24 + \tfrac{1}{4}\nf \big)s - 6 \big(\hD + 4 - \tfrac{3 \nf }{4}\big) n^2 + 24 n^2 \ln n^2 \\
&\qquad\qquad \ \  \te \te +6 n^2 \big[(\hD- \frac{\nf }{2}) n^2 + 2 s (1- \tfrac{\nf }{16}) \big] Q_n(s)  + 12 n^2 s(1- \frac{\nf }{16}) Q_n(-s)  \Big]~.
\end{aligned}
\end{equation}
In the special dimension $D=11$ where  $\hD=8$ and $\nf =16$,
 these  amplitudes simplify to 
 \begin{equation} \label{68}
\begin{aligned}
\hat{A}_n^{(1)} &= \tfrac{1}{16 \pi}  s^2 \big(s + 2 n^2 \ln n^2 \big)~, \\
\hat{B}_n^{(1)} &= \tfrac{1}{32 \pi} s^2 \Big[ -8 n^2 + s^2 Q_n(-s) + s^2 Q_n(s) \Big]~, \\
\hat{C}_n^{(1)} &= \tfrac{1}{16 \pi} s^2\big(-s + 2 n^2 \ln n^2 \big)~.
\end{aligned}
\end{equation}
To restore the  radius $\rr$ dependence   we need to  do the replacement $n \to {n\ov \rr}$,  getting  
\begin{align} 
&\hat{A}_n^{(1)} = \tfrac{1}{16 \pi}  s^2 \Big[ s + 2\rr^{-2}  n^2 \ln (\rr^{-2}  n^2) \Big] ~, \qquad \
 \hat{C}_n^{(1)} = \tfrac{1}{16 \pi}  s^2 \Big[ - s + 2\rr^{-2}  n^2 \ln (\rr^{-2}  n^2) \Big]    \, , \la{688}  \\
&\qquad \qquad \hat{B}_n^{(1)} = \tfrac{1}{32 \pi} s^2 \Big[ -8 \rr^{-2} n^2 +  s^2 \,   Q_n(-\rr^2 s) +  s^2\, Q_n(\rr^2 s)  \Big]~.\la{6688}
\end{align}
Finally, we need  to sum over  the mode number $n\in (-\infty, \infty)$  using as in~\ci{Seibold:2023zkz}
  the standard  Riemann $\zeta$-function relations (here $k$ is a  positive integer)~\foot{Note that since here   we  have
  a  flat 3d space  with  an  $S^1$  factor   the Riemann $\zeta$-function  
 is  the same as the spectral $\zeta$-function    for the 2nd-order operator  in  the compact  direction.}
\begin{equation}
\begin{aligned}\la{69}
&\sum_{n=-\infty }^\infty 1 =1 + 2 \zeta(0) = 0 ~, \quad\qquad \ \ \  \  \sum_{n=1}^\infty n^{2k} = \zeta(-2k) = 0~, \quad \\
&\qquad \qquad  \sum_{n=1}^\infty n^2 \ln n^2 = -2 \zeta'(-2)=   \tfrac{1}{2 \pi^2} \zeta(3) ~. 
\end{aligned}
\end{equation}
Starting  with the UV divergent contributions, 
we observe that   while   the $stu$  terms in the  first lines  in~\rf{61}  and~\rf{65} vanish  already   due to the 
   2d kinematics, 
 being   $n$-independent   they in any case  do not contribute  as $\sum^\infty_{n=-\infty}1 =0$. 
The remaining terms  in~\rf{61}  and~\rf{65} 
  are proportional to $n^2$  and thus   also   vanish  after the   summation  since  $\zeta(-2)=0$. 
  Thus   the total  1-loop scattering amplitude  is  finite  with  the coefficients given by the sums of~\rf{68} 
\begin{equation}
\hat{A}^{(1)} =  \sum_{n=-\infty}^\infty \hat{A}^{(1)}_n~, \qquad
\hat{B}^{(1)} =  \sum_{n=-\infty}^\infty \hat{B}^{(1)}_n~,\qquad 
\hat{C}^{(1)} =  \sum_{n=-\infty}^\infty \hat{C}^{(1)}_n~.\la{70}
\end{equation}
As a result, we find that 
\begin{align}
&\hat{A}^{(1)} = \hat{C}^{(1)}  = \tfrac{1}{8 \pi^3}  \zeta(3)\, \rr^{-2} \, s^2~, \ \la{71} \\
& \hat{B}^{(1)} = \tfrac{1}{32 \pi}  s^4 \sum_{n=-\infty}^{\infty} \Big[ Q_n(-\rr^2 s) + Q_n(\rr^2 s) \Big]~.\la{72}
\end{align}
As discussed in   the bosonic  membrane case  in~\ci{Seibold:2023zkz},  for 
the  theory on  $\mathbb R^{1,1} \times S^1$    there  is also a   non-zero 
 contribution  coming from the tadpole diagram  in Fig.~\ref{f3} that  cancels  the $\zeta(3)$  terms in 
  $ \hat{A}^{(1)} $  and $ \hat{C}^{(1)} $
   in~\rf{71}. As shown in Appendix~\ref{C}, the same also  happens here      so that  in total 
  \begin{equation} \la{711} 
  \hat{A}^{(1)} = \hat{C}^{(1)}  = 0 ~.
   \end{equation}
   This is 
     in agreement with the  observation 
    that restricting the amplitude~\rf{333},\rf{51}
   on  $\mathbb R^{1,2}$  to the  planar 2d kinematics one finds that  the  annihilation and  reflection amplitudes 
   vanish, i.e. the scattering becomes diagonal.

Observing that the  function $Q_n(s)$ in~\rf{64}  has the following asymptotic behaviour 
\begin{equation}\te 
Q_n (s)\Big|_{n\to 0}  = -\frac{2}{s}\, \ln \frac{n^2}{s}  +\OO(s^{-1} n^2)\ , \qquad \ \ 
Q_n(s)\Big|_{n\to \infty} = \frac{1}{n^2} -\frac{s}{6 n^4} + \OO({s^2\ov n^6}) \,,\la{73}
\end{equation}
we may   separate   the $n=0$ contribution  in~\rf{72} representing it as  
\begin{align}
&\hat{B}^{(1)} = \tfrac{i}{16}s^3  -  \tfrac{1}{16 \pi} s^3 \sum_{n=1}^\infty \Big[\bQ_n(-\rr^2 s )-  \bQ_n(\rr^2 s)\Big]~, \la{74}\\
&\bQ_n(x) \equiv  x\, Q_n(x) =   - \tfrac{2}{ \sqrt{1+\frac{4 n^2}{x}}} \ln   \tfrac{ \sqrt{1+\frac{4 n^2}{x}}-1}{\sqrt{1+\frac{4 n^2}{x}}+1}~.
\la{744}
\end{align}

\subsection{$\rr\to 0$ and $\rr\to \infty$ limits}

As is well known,   the double dimensional reduction of the  classical 
 $D=11$  supermembrane   action   gives the   type IIA    GS superstring
action~\ci{Duff:1987bx,Achucarro:1989dd}. 
Then   by taking the  $\rr\to 0$   limit  of  the   $\mathbb R^{1,1}\times S^1 $  supermembrane 
S-matrix for the 2d massless modes  we should 
  get   the S-matrix  on the   superstring with $\mathbb R^{1,1} $ world sheet
  (the $\rr\to 0$   limit    should decouple  massive 2d mode  contributions in the loops). 
  From~\rf{74},\rf{711}   we find
 \begin{equation} \la{76}
 \hat{A}^{(1)}\Big|_{\rr\to 0} = \hat{C}^{(1)}\Big|_{\rr\to 0} =0 \ , 
 \qquad \qquad \hat{B}^{(1)}\Big|_{\rr\to 0} = \hat{B}^{(1)}_0 = \tfrac{i}{16} s^3 \ ,  
 \end{equation}
  which is indeed  the expected  $D=10$ superstring theory result (cf.~\ci{Cooper:2014noa,Mohsen:2016lch}).
 In the superstring limit  we need to fix  the  string tension 
 $T_1= 2 \pi \rr\, T_2= {1\ov 2\pi \a'}$   and the scattering amplitude takes the form (cf.~\rf{333}) 
 $\hat {\cal M}=  \delta^{I_1I_3} \delta^{I_2 I_4}  (\ha s^2   + {i\ov 16  T_1}    s^3 + ... ) $
 which is the same as in bosonic case. 
 We are ignoring extra overall factors  like $2\pi \rr$ associated  with  the momentum delta-function, see~\ci{Seibold:2023zkz}.

 Let us note that  the   subleading in small  11d  radius $\rr$ terms   in the expansion of the  amplitude \rf{74} 
 have the interpretation of the leading  in small $s$  or large string tension  terms  at each order 
 in string coupling $g_s$  expansion. 
 As in the examples discussed in~\ci{Giombi:2023vzu,Beccaria:2023ujc,Beccaria:2023sph}
 the 1-loop M2  brane correction sums  up all leading large tension  terms at each order in $g_s$ expansion (i.e.  $\sum_{n=1}^\infty c_n ({g^2_s\ov T_1})^n$)   that  should effectively  correspond to  a sum of   parts of string loop corrections with small  handles inserted on a  plane.\foot{One 
 may conjecture  that the effect of handle insertions may be 
 represented by an effective  non-local 2d  theory (cf. \ci{Tseytlin:1990vf, Skliros:2019bqr}).
 In the present case  integrating 
 out massive modes (with  inverse masses 
 proportional to $R_{11}$ or $g_s$)
  leads  to an effective  action for massless (string-level) excitations that 
 should be playing this role.}   
 Here\foot{We follow the standard notation (see e.g. Appendix A in \ci{Beccaria:2023hhi}) expressing  $T_2= {1\ov (2\pi)^2\ell_{pl}^3}$ or $\ell_{pl}$ in terms of $\rr$,  $g_s$  and $\a'$.}
 $\rr = \sqrt{2\pi \a' } g_s$  and expanding \rf{74} in small $\rr^2 s$ we  capture the leading 
lowest  power of $\a'  s$ at each order in $g_s$
 \be \la{4477}
 \hat{B}^{(1)} = \te \frac{1}{16}s^3\Big( i    +  {2\pi \ov 3} \a' s  g^2_s   +  {8\pi^8  \ov 14175} \a'^3  s^3 g^6_s + ... \Big) \ ,
 \ee
where we used that  $\zeta(2) = {\pi^2\ov 6}, \ \zeta(6) = {\pi^6\ov 945}$, etc.

To analyse the  ``decompactification"  $\rr \rightarrow \infty$ limit of the 1-loop 
 amplitude~\rf{72} or~\rf{74} we may use  the following relation (see also~\ci{Seibold:2023zkz})
\begin{align}
&\sum_{n=-\infty}^{\infty} Q_n(s) = \int_0^1 dx \, \Big[ \frac{\pi}{\sqrt{\Delta_{-s} (x) }} + \frac{2 \pi}{\sqrt{\Delta_{-s}(x)}} \LL\big(e^{-2 \pi \sqrt{\Delta_{-s}(x)}}\big) \Big]  ~,\la{77}\\
&\qquad  \Delta_{s} (x)  = - x ( 1-x) s - i \ve \ , \ \ \ \  \qquad \LL(y) \equiv \text{Li}_0(y) = \frac{y}{1-y}~. \la{78}
\end{align}
Starting with~\rf{74}, i.e.  introducing  the dependence on $\rr$  in $Q_n$ by $s\to \rr^2 s$   and thus 
 $\Delta_s \to \rr^2 \Delta_s$   we   get~\foot{We used  that $\sqrt{\Delta_s} \rightarrow \sqrt{\Delta_s-i \varepsilon} = \sqrt{-x(1-x) s - i \varepsilon} = -i \sqrt{|\Delta_s|} + \tilde{\varepsilon}$ where $ \tilde{\varepsilon}>0.$}
\begin{equation}
\LL\big(e^{-2 \pi \rr \sqrt{\Delta_{-s}}}\big)  \Big|_{\rr \to \infty} 
= 0~, \qquad 
\LL\big(e^{-2 \pi \rr \sqrt{\Delta_s}}\big) \Big|_{\rr \to \infty}
 = 
  \LL\big(e^{2 \pi i \rr \sqrt{\Delta_{-s}}\, -2 \pi \rr \tilde{\epsilon}}\big) \Big|_{\rr \to \infty}= 0~. \la{79}
\end{equation}
Thus   from~\rf{74}we get that 
\begin{equation}\la{80}
\hat{B}^{(1)} \Big|_{\rr \to \infty} = \tfrac{\pi }{32} s^2 \Big[ (-s)^{3/2}+(s)^{3/2}\Big]~.
\end{equation}
Since~\rf{711}  implies that   $\hat{A}^{(1)} \Big|_{\rr \to \infty} = \hat{C}^{(1)} \Big|_{\rr \to \infty}=0$,  
 the  total 1-loop amplitude  in this limit  corresponds indeed 
 to 
 the restriction to  the  2d kinematics ($t=0,\, u=-s$)  of the  non-compact   $\mathbb R^{1,2}$  supermembrane 
 scattering amplitude  in~\rf{51}.~\foot{More precisely,  to compare the planar scattering limit of~\rf{51} 
 and~\rf{80}  we need to include the momentum delta-function factors, i.e. 
 to  take into account  an extra $2\pi \rr$   appearing  together with the mode number  delta-function
 (see~\ci{Seibold:2023zkz}  for details about normalisations).}

\section{Relation to one-loop scattering   amplitude on   $D=10$ superstring}
\label{s5}

Since  the double dimensional reduction of the supermembrane  action     gives the   one for the    GS superstring, 
we   should  also   get the S-matrix  on  the later   by simply truncating the   supermembrane 
expressions to  massless $n=0$ level  at each step of the calculation. 
Explicitly, 
the  dimensional reduction of the membrane action~\rf{13}--\rf{15} along $X^2=\sigma^2$ gives~\foot{In this section 
we rename  the  coordinates so that 
$\m,\nu=0,1$  and $M,N=0,1,3,..,D-1$.  To recall, 
in the    standard  notation  is denoted as  $X^{10}$   and then  $\Gamma_2$   is replaced by  $\Gamma_{10}$. }
\begin{equation}\la{5.1}
\begin{aligned}
&S = S_1 + S_2~, \qquad \qquad S_1 = - T_1 \int d^2 \sigma \sqrt{-\det \hat g_{\mu \nu}}~, \\ &S_2 = - T_1 \int d^2 \sigma \, i \, \varepsilon^{\mu \nu}\,  \bar{\theta} \Gamma_{M} \Gamma_2 \partial_{\mu} \theta \big( \Pi_{\nu}^{M} + \tfrac{i}{2} \bar{\theta} \Gamma^{M} \partial_{\nu} \theta \big)~,  \\
&\hat g_{\mu \nu} = \eta_{MN} \Pi_\mu^M \Pi_\nu^N~, \qquad \Pi_\mu^M = \partial_\mu X^M - i \bar{\theta} \Gamma^M \partial_\mu \theta~.
\end{aligned}
\end{equation}
Here $ \varepsilon^{\mu \nu} = \epsilon^{\mu \nu 2}$  and $\theta$ is the same fermionic 
spinor variable   as in the  supermembrane action 
(e.g.  an 11d  Majorana  spinor    or  a pair of  opposite-chirality 10d MW  spinors). 
The  supersymmetric and $\k$-symmetric superstring action then exists in dimensions $\rD=D-1=3,4,6,10$. 
Expanding near the infinite string vacuum corresponds to  choosing the  analogous static   gauge  and $\k$-symmetry gauge 
as in the membrane case 
(cf.~\rf{21},\rf{28}; \  $I=3,\dots,D-1$)
\begin{equation}\la{5.2}
X^M = (\sigma^\mu, X^I)~,  \qquad \qquad \ \mathcal P_+ \theta =0 \ , \ \ \  \ \ \  \mathcal P_\pm  = \tfrac{1}{2} (1\pm \Gamma^\star)\ , \ \ \ \ \ 
\Gamma^\star = \G^{01}\G^2 \ . \end{equation}
This  leads to the  same expressions as  in~\rf{54},\rf{57} and~\rf{58} restricted to the massless   level, i.e.
  with all mode  numbers $n_i=0$ 
\begin{equation}\la{5.3}
\begin{aligned}
\mathcal L_2 &= -\tfrac{1}{2} \partial_\mu X^I \partial^\mu X^I + i \bar{\theta} \mathcal P_- \Gamma^\mu \partial_\mu \theta~, \qquad\qquad   
\mathcal L_3 =0~, \\
\mathcal L_4 &=-\tfrac{1}{8} \partial_\mu X^I \partial^\mu X^I \partial_\nu X^J \partial^\nu X^J +\tfrac{1}{4} \partial_\mu X^I \partial_\nu X^I \partial^\nu X^J \partial^\mu X^J \\
&\qquad + \tfrac{i}{4} \partial_\mu X^I \partial^\mu X^I \bar{\theta} \mathcal P_- \Gamma^\nu  \partial_\nu \theta - \tfrac{i}{2} \partial_\mu X^I \partial^\nu X^I \bar{\theta} \mathcal P_- \Gamma^\mu   \partial_\nu\theta ~.
\end{aligned}
\end{equation}
The corresponding  
 tree-level scattering   amplitude for 4  massless   scalars $X^I$  in $\rD=10$  superstring 
 is the same as in the bosonic string   while  the 1-loop 
 amplitude    can  then  be  found    by  simply
   setting $n=0$   in~\rf{68}.\foot{Note that 
    as follows from \rf{61}  and \rf{65}  before specifying  to 2d kinematics  (i.e. setting $stu=0$)
the superstring amplitude   contains the following divergent term:
$\hat {\cal M}^{(1)} 
=-  \te \frac{1}{96 \pi\, \epsilon} \, q\, 
  s t u \, \big( \delta^{I_1 I_2} \delta^{I_3 I_4} + \delta^{I_1 I_3} \delta^{I_2 I_4} + \delta^{I_1 I_4} \delta^{I_2 I_3} \big) $. 
Here $q= \hat D - 6 + {1\ov 4} \nf = {\rm D} -8 +4 =6$  where we  specified to the superstring  values
 $\rm D=10$ and $ \nf=16$. 
 This is   the same as the   coefficient  of the 
 1-loop UV   divergence $\sim  \int d^2\s\,  \sqrt{-g}\, R$   in the effective action of the 
  GS string on a general  curved 2d background  \ci{Forste:1999qn,Drukker:2000ep,Giombi:2020mhz}:
  $\Gamma_\infty =-{1\ov 4\pi }  \log \Lambda \int  d^2\s\,  \sqrt{-g}\, b_2$ where $b_2= {1\ov 6} q R$  and 
  $\log  \Lambda$ stands for ${1\ov 2 \epsilon} $.
  In general,  for an operator $-\nabla^2 + U$   one has  $b_2 = \tr ( {1\ov 6} R - U)$.
  In the case of the GS   string  in conformal gauge   the 1-loop expression contains  the contribution of 
  $\rm D=10$ massless   scalars and also  the  bosonic ghost  one  with $-\nabla^2_{\mu\nu} - \ha R g_{\mu\nu}$ 
    or, equivalently,  in the static gauge, of $\rm D-1$ massless   scalars  and one scalar  with mass term  $U=R$,  
   giving $b^{(b)}_2= ({\rm D}-2) {1\ov 6 } R  -   R=({\rm D}-8)   {1\ov 6 } R   $.  In addition, there is also  the fermionic  contribution of   $\ha  \nf=8$  squared    Dirac operators  $-\hat \nabla^2 + {1\ov 4} R$ 
   giving $b^{(f)}_2 = -{1\ov 2}  \nf \times (  {1\ov 6 } R  - {1\ov 4}   R)=  {1\ov 4}  \nf \,      {1\ov 6 } R    $. In total, 
    $b_2=  b^{(b)}_2 + b^{(f)}_2 =  R$ which agrees with \ci{Giombi:2020mhz}  and 
    corresponds to $q={\rm D}-8  +  {1\ov 4}  \nf=  6$
     (this  corrects  a  misprint in eq. (3.28) 
   in  \ci{Drukker:2000ep}; the  above  expression for $q$  follows also  from (6.50)-(6.54) there). 
} 
 In general, the   finite part of the fermionic  loop contribution is 
 $A^{(1)}_F = -\frac{1}{192 \pi} \frac{\nf}{12} \,\big[ 3 s^3 +4s  t u - 6s t u \log ( -{s\ov \mu^2})
 \big] $  and similarly (by interchanging $s,t,u$)  for $B^{(1)}_F $  and $C^{(1)}_F$.

Surprisingly, this gives a different result  for $\hat {A}^{(1)}$ and  $\hat {C}^{(1)}$   than in~\rf{76}:    
\begin{equation} \la{81} 
\hat {A}^{(1)}_0  =- \hat{C}^{(1)}_0  = \tfrac{1}{16\pi } s^3  \ , 
 \qquad \qquad  \hat{B}^{(1)}_0 = \tfrac{i}{16} s^3 \ .  
 \end{equation}
  The non-vanishing  of  $\hat {A}^{(1)}$  and $\hat {C}^{(1)}$   following directly   from~\rf{5.3}  is  in  an apparent 
 conflict with the expected  elastic scattering in the critical superstring case (cf.~\ci{Cooper:2014noa,Mohsen:2016lch}).

     This ``anomaly'', i.e.  that the $\rr\to 0$ limit  of the membrane theory  does not commute with  quantization/regularization, 
   follows   from  
    the   fact that   the leading  $s^3 $ terms in $\hat {A}^{(1)}_n $  and $\hat {C}^{(1)}_n $  in~\rf{68}
   do not  depend on $n$  and thus on $\rr$   
   so that they   {\it do not }  decouple
    in the $\rr\to 0$ limit. 
   The vanishing of $\hat{A}^{(1)}$ and $\hat{C}^{(1)}$  in~\rf{76}    is due  effectively  to 
   the contribution of   all non-zero $n$   modes  that cancels  the  massless $n=0$  contribution in~\rf{81}  
  as a consequence of the   $1 + 2 \zeta(0)=0$   regularization relation in~\rf{69}. 

To understand the reason for this anomaly  let us recall also  the  analog of~\rf{81}  for the  1-loop amplitude in a general 
 superstring (bosonic plus fermionic loop)  theory  with dimension $\rD=D-1$   that follows from setting $n=0$ in~\rf{67}:
   \begin{align}   & \hat {A}^{(1)}_0  =- \hat{C}^{(1)}_0  = -\tfrac{1}{192\pi } \, \bD\, s^3\ ,   
 \qquad \qquad  \hat{B}^{(1)}_0 = \tfrac{i}{16} s^3 \ ,  \qquad  \la{82}\\
 & \bD= \hD - 24 + \tfrac{1}{4}\nf = \tfrac{3}{2} \hD -24 
 \ , \qquad \hat D = D-3 = \rD-2 \ . \la{83}\end{align}
  Then~\rf{81} corresponds to $\hat D= 10-2=8, \  \nf=16$, i.e. ${\rm b}\big|_{{\rD=10}}=-12$.  
   
 Note that the  coefficient $\bD$ is also the one  that appears in the computation of the conformal anomaly 
  of the GS   superstring  with  a naive  treatment of  $\theta$'s  as 2d fermions rather  than 2d scalars. 
   In fact,  each GS   fermion contributes  to the conformal anomaly  4 times greater than one  2d Majorana fermion
    (see~\ci{Kallosh:1987hkh,Wiegmann:1989md,Lechner:1995yi}  and  a discussion in~\ci{Drukker:2000ep})\foot{\la{foo}The reason for this  is that 
      the  kinetic term for $\theta$ which is a 2d scalar  in the GS  Lagrangian  in~\rf{5.1}   has the structure 
     $\sqrt{-g}\, g^{\mu\nu} \del_\mu X^M  \bar \theta \Gamma_M \del_\nu \theta$ 
     which is to be compared to $\sqrt{-g}\,  e^\nu_a  \bar \psi \Gamma^a \del_\nu \psi$  for a 2d spinor.
     In the conformal gauge $g_{\mu\nu} =e^{2\rho} \eta_{\mu\nu},
\ e^a_\mu = e^{\rho }\d^a_\mu $ that effectively results in
the replacement of $\rho$ by $2\rho$ in the 
conformal anomaly
term $S_P= {\bD\ov 24\pi }  \int  d^2\s \, \del \rho \bar \del \rho$ coming from  a 2-d
spinor, giving   4  times bigger a result.   One can give  an 
equivalent argument  separating the 2d  conformal factor 
dependence into  a  Jacobian  originating from    a rotation
of  the GS   fermion    (see~\ci{Kavalov:1986nx,Langouche:1987mx,Wiegmann:1989md} 
and  Appendix C in~\ci{Drukker:2000ep}).
 }
  so that 
    the  total conformal   anomaly  coefficient is 
     actually $\bD_c=\hD - 24 +4 \times  \tfrac{1}{4}\nf = 3 \hD -24 =   3( \rD- 10)$  
      so that  it vanishes  for $ \rD=10$.
  
  The   direct expansion of the GS action~\rf{5.1}   in the static gauge with the induced metric 
  $g_{ \m \nu}=\eta_{\mu\nu} +   \del_\mu X^I \del_\nu X^I$    treats  $\theta$  in~\rf{5.3} 
  as a collection  of $\ha \nf$    2d Majorana  fermions. This   suggests  that we are effectively 
    missing the   contribution  of a local  counterterm  
  that cancels  the  $s^3$ terms  in ${A}^{(1)}$ and ${C}^{(1)}$  in~\rf{81},\rf{82}.

To motivate this explanation 
 let  us  recall  that in the bosonic  Nambu string  in $\rD=\hD +2 $   dimensions  the scattering amplitude for the massless 
  bosons  in the static  gauge   has the   1-loop coefficients~\ci{Dubovsky:2012sh,Seibold:2023zkz}
 that are the special case of $\nf=0$  of~\rf{82}, i.e. with $\bD=  \hD-24= \rD-26$.
Thus in the   critical   dimension  $\rD=26$ or   $\hD=24$   the 1-loop amplitude 
 contains  only the $B^{(1)}$ term, i.e.  like the tree-level  amplitude
   (with  $\hat A^{(0)}= \hat C^{(0)}=0, \  \hat B^{(0)}=  \ha s^2$,  cf.~\rf{1},\rf{3})  it is proportional  to the identity and  satisfies the Yang-Baxter equation. 

The fact that the annihilation and  reflection amplitudes $\hat A^{(1)} $ and $\hat  C^{(1)} $  in~\rf{82}  are real and  analytic in $s$ 
   for any  value of $\bD$   suggests   that  it should be possible to   set them  to zero 
   and thus    restore the elastic structure of the scattering amplitude 
       by  adding a  contribution of  a local   6-derivative    counterterm.\foot{Another reason for that  is 
       that for  generic  $\hat D$   there is a  relation  between  the  static-gauge Nambu   action   and the   $T\bar{T}$ deformation
   of the  theory of $\hat D$  free scalars~\ci{Cavaglia:2016oda}  (cf.~also~\ci{Baggio:2018rpv}).  }  
       The relevant  one is~\ci{Dubovsky:2012sh,Aharony:2013ipa}
\begin{equation}
\label{237}
 \Delta S_4 = - 2b \int d^2\s  \,  \big( \partial^\mu  X^I\partial_{ \mu}  \partial_ \nu X^I\big)^2  
   \  .
\end{equation}
Indeed, it   leads  to the following tree-level  scattering  amplitudes  (assuming  $t=0$  or $u=-s$)
\begin{equation}
\Delta A =- \Delta C =    b\,  s^3\,  ,\qquad  \qquad\ \  \Delta B =  0    \ . \la{238}
\end{equation}
Thus   adding~\rf{237}  one can   cancel     the non-zero values of  $\hat A^{(1)}$ and  $ \hat  C^{(1)}$ in~\rf{82}   by  choosing 
$b = \tfrac{1}{192\pi} \bD $.

The counterterm~\rf{237}  originates 
 from the Polyakov conformal anomaly term~\ci{Polyakov:1981rd}
 \begin{equation} \la{84}
 S_P=  2 b  \int R^{(2)} {\nabla}^{-2} R^{(2)}= \Delta S_4 +  \OO(X^6) 
  \ , \ \ \ \ \ \ \ \  b = \tfrac{1}{192\pi} \bD \ ,  
  \end{equation}
after one evaluates~\rf{84}  on the induced metric $g_{ \m \nu}=\eta_{\mu\nu} + \del_\mu X^I \del_\nu X^I$ 
 and expands  in powers of 
derivatives of the transverse coordinates
(cf.~\ci{Polchinski:1991ax,Aharony:2013ipa}).~\foot{One may use that  
$R^{(2)} = \del_\m \del_\nu h^{\m\nu} - \del^2 h + ...$   where $ h_{\m\nu}= \del_\mu X^I \del_\nu X^I$  and    ignore terms 
 proportional  to the equation of motion factor  $\del^\m\del_\m X^I$ 
 as we are interested in the  on-shell amplitude following from this  vertex.}    
 
 {\com  In the bosonic   string  case  in non-critical  dimension adding the  counterterm \rf{237}  to preserve integrability  would break the Poincare invariance of the original theory  (fixing  static gauge  is  effectively consistent with the Poincare invariance  only assuming  the   cancellation of  the conformal anomaly).}
 
In the present  GS string  context, the   direct  expansion of the action~\rf{5.1}  in the static gauge 
should  also be  supplemented by the term~\rf{84} or~\rf{237}  with $\bD$ given by~\rf{83}, 
 i.e.   with $\bD=-12$  in the only consistent case of  the $\rD=10$  critical superstring.\foot{In general, this 
 term  is to be accompanied   by a logarithmic divergence  $- {1\ov 24\pi}    \bD \log \Lambda \int d^2\s\, \sqrt{-g} R^{(2)}$
 which contributes in the case of non-trivial world-sheet topology 
  and  is needed (along with other similar contributions from the path integral measure) 
    to make the full superstring partition function UV finite and  conformal anomaly free 
    in  $\rD=10$ as discussed  in~\ci{Drukker:2000ep} and~\ci{Giombi:2020mhz}.}
   {\com The   need for the  extra term \rf{84}   is directly  related  to the fact   that it is also required  for the conformal anomaly cancellation  in the $\rD=10$  GS string theory which is non-trivial  and depends on precise definition of the path integral measure  (cf. footnote \ref{foo}).}

Remarkably, defining the quantum  superstring as  the $\rr\to 0$ limit of the  $S^1$-compactified  $D=11$ 
 supermembrane automatically produces 
 this counterterm  from the sum over all non-zero mass terms
   regularized   with the $\zeta$-function as in~\rf{69}.\foot{This  procedure also  produces a 
   UV finite result   in  non-trivial cases like the ones discussed in~\ci{Giombi:2023vzu,Beccaria:2023ujc,Beccaria:2023sph}.}

\iffa
naive  treatment of GS string requires  Jacobian in the measure.
it accounts for cancellation of UV div and conf anomaly  and consistent S-matrix. 
in correct treatment of GS  near long string this jacobian comes from rotating from GS to 2d fermion representation. 
In semiclassical expansion for partition function  this matters only when    induced   metric is non-trivial (not flat); for correlators or S-matrix  this matters. 

Defining GS string as limit of membrane automatically  produces consistent result assuming 
zeta-function   regularization (that is natural one here --   needed for consistency with 3d treatment where extra -- 3d Lorentz -- symmetry is manifest -- no log divergences).  Jacobian here comes   effectively from contribution of all massive modes on $S^1$. 
\fi

\section*{Acknowledgements}
We are  grateful  to  S.~Dubovsky,  H. Jiang, R. Metsaev, R.~Rattazzi,   
A.~Sfondrini and A.~Vainshtein   for  discussions.
We also thank S.~Giombi and R.~Roiban for useful  suggestions and comments on the draft. 
F.S.~would like to acknowledge  the hospitality of New York University and the Simons Center for Geometry and Physics, Stony Brook University, where part of this work was done. F.S. also thanks the MATRIX Institute in Creswick \& Melbourne, Australia, for support through a MATRIX-Simons Young Scholarship in conjunction with the MATRIX program ``New Deformations of Quantum Field and Gravity Theories''. 
FS  is supported  by  the European Union  Horizon 2020 research and innovation programme under the Marie Sklodowska-Curie grant agreement number 101027251.   
Some   of this work was done while AAT was attending  KITP program ``What is String Theory?"   in March 2024 where his research  was supported in part by grant NSF PHY-2309135
 to the Kavli Institute for Theoretical Physics (KITP). 
This work was also    supported by the STFC grant ST/T000791/1.

\small 
\appendix

\section{Conventions and spinor traces}
\label{A}
For a $2 \rightarrow 2$  massless scattering process  we define the Mandelstam variables as 
 \begin{equation} \la{a1}
 \begin{aligned}
 s &= -(p_1+p_2)^2 = -(p_3+p_4)^2 = -2 p_1 \cdot p_2 = -2 p_3 \cdot p_4~, \\
 t &= -(p_1-p_3)^2 = -(p_2-p_4)^2 = 2 p_1 \cdot p_3 = 2 p_2 \cdot p_4~, \\
 u &= -(p_1-p_4)^2 = -(p_2-p_3)^2 = 2 p_1 \cdot p_4 = 2 p_2 \cdot p_3~, \qquad  s + t + u  =0~.
 \end{aligned} 
 \end{equation}
 We  assume   that $\eta_{\m\nu}={\rm diag}(-1,1,1)$, $ \epsilon^{012} =-1$  and 
$
\epsilon_{\mu \alpha \beta} \epsilon^{\mu \gamma \delta} = \delta_\alpha^\delta \delta_\beta^\gamma - \delta_\alpha^\gamma \delta_\beta^\delta~.
$

To recall, in the supermembrane actions in allowed dimensions $D=11,7,5,4$  the fermionic variable 
$\theta$ is represented by  the  following  target-space spinors  with $2\nf$ real components 
\ci{Achucarro:1987nc}
\begin{equation} \la{a7}
\begin{aligned}
D&=11:& \qquad &\text{ Majorana} &\rightarrow\qquad  \nf  &= 16 
~, \\
D&=7:& \qquad  &\text{ Symplectic Majorana} &\rightarrow\qquad  \nf  &= 8 
~, \\
D&=5:&\qquad &\text{ Symplectic Majorana} &\rightarrow\qquad  \nf  &= 4 
~, \\
D&=4:&\qquad &\text{ Majorana or Weyl} &\rightarrow\qquad  \nf  &= 2  
~.
\end{aligned}
\end{equation}
Fixing $\k$-symmetry gauge leaves  one  with $\nf$ real  fermionic components. 
The  number of  real physical  3d scalars 
 $\hD=D-3$   then matches the number of the 3d  fermionic degrees of freedom $\ha \nf$ 
in each of these  dimensions, i.e.
\begin{equation} \la{a8} 
\half \nf = \hD \ . 
\end{equation}
To compute  traces over the spinor indices we use the following relations (here $\mathcal P_-$  is the 
projector implementing $\k$-symmetry gauge fixing in~\rf{28},\rf{29})
\begin{align} 
&\Tr[\mathcal P_-\Gamma_{\mu} \Gamma_{\nu} \Gamma_{\rho} \Gamma_{\sigma}] = \left(\eta_{\mu \nu} \eta_{\rho \sigma} - \eta_{\mu \rho} \eta_{\nu \sigma} + \eta_{\mu \sigma} \eta_{\nu \rho} \right) \Tr[\mathcal P_-]~,
\la{a2} \\
&\Tr[\mathcal P _-\Gamma_{\mu_1} \dots \Gamma_{\mu_n} \Gamma^{IJ}]=0~,\la{a3} \\
&
\Tr[\mathcal P _-\Gamma_{\mu} \Gamma_\nu \Gamma^{I_1 I_2} \Gamma^{I_3 I_4}] = \eta_{\mu \nu} (\eta^{I_1 I_4} \eta^{I_2 I_3} - \eta^{I_1 I_3} \eta^{I_2 I_4}) \Tr[\mathcal P_-]~,\la{a4} \\ 
& \Tr[\mathcal P_-] =\half \nu_{_D}\equiv \nf \,, \ \ \ \ \ \ \ \ \  \Tr[ 1 ]= \nu_{_D}= 2^{[ \frac{D}{2}]} \, . \la{a5}  
\end{align}
Again, $\nf$ is  the  number of  {real} spinor components   {after} $\k$-symmetry gauge fixing.
Note   that  when computing   fermion loop  for a Majorana  spinor  we need to add an extra coefficient $\ha$ 
(compared   to  the Dirac spinor case) 
   in front of  the spinor trace  as in~\rf{44}, i.e.  effectively 
 $ \Tr[\mathcal P_-...]\to \half  \Tr[\mathcal P_-...]$ or $ \Tr[\mathcal P_-]  \to \half \nf$. 
 This was   assumed in~\rf{48}. 

\iffa 
In general dimension $D$  we then  have 
\begin{equation} \la{6}
\begin{aligned}
 \nf  &= 2^{[ \frac{D}{2}]}&\qquad &\text{for Dirac or Symplectic Majorana spinors}~, \\
 \nf  &= \tfrac{1}{2}\,  2^{[\frac{D}{2}]}&\qquad &\text{for Majorana or Weyl spinors}~, \\
  \nf  &=\tfrac{1}{4} \, 2^{[\frac{D}{2}]}&\qquad &\text{for Majorana-Weyl spinors}~.
 \end{aligned}
\end{equation}
\fi

\section{Comments on unitarity}
\la{B}

The  expression  for the 1-loop  amplitudes in~\rf{88} or~\rf{111}
  contains  an  imaginary part that should be consistent with  the unitarity of the S-matrix 
(i.e. should   be  related to the square of the of tree-level  amplitude  via cutting).   
To recall, in general,  
 the unitarity $\rS^\dagger \rS=1$  of the S-matrix  $\rS = 1 + i \rT $ implies that 
 $2\,  \text{Im} \rT = \rT^\dagger \rT$. 
 In the case of compactified membrane, 
for the  1-loop   contribution  with the  bosonic mode number $n$ in the loop (corresponding to the bosonic  membrane case) 
  in~\rf{62},\rf{63}  starting with the 4-scalar  tree-level vertex  following from~\rf{57} 
  one finds that 
    the unitarity  condition for the S-matrix implies  the following conditions 
\begin{align}
\Ima [A_{B,n}^{(1)}] &=  \tfrac{1}{16} n^2 s(\hD n^2 - 2 s) \frac{\Theta(s-4 n^2)}{\sqrt{1-\frac{4 n^2}{s}}} ~, \la{b1}\\
\Ima [B_{B,n}^{(1)}] &=  \tfrac{1}{16}s^2  (s-2 n^2) \frac{\Theta(s-4 n^2)}{\sqrt{1-\frac{4 n^2}{s}}} ~, \qquad 
\Ima [C_{B,n}^{(1)}] = \tfrac{1}{8}n^2 s^2\frac{\Theta(s-4 n^2)}{\sqrt{1-\frac{4 n^2}{s}}}~,\la{b2}
\end{align}
where $\Theta$ denotes the Heaviside step function. The agreement with~\rf{62}--\rf{64}  
 follows from the fact that $Q_n$ in~\rf{64} satisfies 
\begin{equation}\la{b3}
\Ima \big[Q_{n \neq 0}(s)\big] = 0~, \qquad \Ima \big[ Q_{n\neq 0}(-s) \big]= \frac{2 \pi}{s} \frac{\Theta(s-4 n^2)}{\sqrt{1-\frac{4 n^2}{s}}}~, \qquad Q_0(s) + Q_0(-s) = \frac{2 i \pi}{s}~.
\end{equation}
A similar  check is  possible for the fermionic contribution to the  loop leading to~\rf{66}. 

One can also  consider  the case of the uncompactified membrane  and  explicitly 
check the compatibility of the 
3d amplitude~\eqref{51} with unitarity. This is best done by writing the amplitude in terms of the Mandelstam variable $s$, related to the center-of-mass energy, as well as the scattering angle $\varphi$  defined as in \rf{9}. The amplitude can be decomposed into partial waves
\begin{equation}
\mathcal M^{(n)}(s,\varphi) = \sum_{l=0}^\infty  \epsilon_l M_l^{(n)}(s) \cos(l \varphi)~, \qquad \int_{- \pi}^{\pi}  d \varphi\, \cos (l \varphi) \cos (k \varphi) = \frac{2\pi}{\epsilon_l} \delta_{l k}~, \qquad l,k \in \mathbb{N}~,
\end{equation}
where $\epsilon_0=1$ and $\epsilon_l=2$ for $l>0$. We suppressed the $SO(\hat{D})$ indices and spinor indices.
Denoting by $\varphi_{iv}$ the scattering angle between the incoming particles $(p_1,p_2)$ and the intermediate particles, as well as $\varphi_{vf}$ the scattering angle between the intermediate (virtual) 
particles and the outgoing states $(p_3,p_4)$, one has the relation $\varphi = \varphi_{iv}+ \varphi_{vf}$. One can then write $\varphi_{iv} = \frac{\varphi}{2} + \phi$ and $\varphi_{vf} = \frac{\varphi}{2} - \phi$, and unitarity (to one-loop order) imposes
\begin{equation}
\begin{aligned}
&\sum_{l=0}^\infty \epsilon_l \Ima[M_{l;  I_1 I_2 \rightarrow I_3 I_4}^{(1)}(s)] \cos(l \varphi)  \\
&= \frac{1}{16 s^{1/2}} \int_{-\pi}^\pi \frac{d \phi}{2 \pi} \sum_{\alpha}   \sum_{l_1,l_2=0}^{\infty} \epsilon_{l_1} \epsilon_{l_2} M_{l_1;  I_1 I_2 \rightarrow \{\alpha\}}^{(0)}(s) [M_{l_2 ;  I_3 I_4 \rightarrow \{\alpha\}}^{(0)}(s)]^\star \cos( l_1 \varphi_{iv})  \cos( l_2 \varphi_{vf}) \\
&= \frac{1}{16 s^{1/2}}  \sum_{l=0}^{\infty} \sum_{\alpha} \epsilon_l  M_{l;  I_1 I_2 \rightarrow \{\alpha\}}^{(0)}(s) [M_{l ;  I_3 I_4 \rightarrow \{\alpha\}}^{(0)}(s)]^\star \cos( l \varphi)~,
\end{aligned}
\end{equation}
where $\{\alpha\}$ collectively denotes all the indices of the intermediate states, namely the $SO(\hat{D})$ indices of the intermediate bosons and the spinor indices of the intermediate fermions. Therefore, we must have
\begin{equation}
 \Ima[M_{l;  I_1 I_2 \rightarrow I_3 I_4}^{(1)}(s)] = \frac{1}{16 s^{1/2}} \sum_{\alpha}  M_{l;  I_1 I_2 \rightarrow \{\alpha\}}^{(0)}(s) [M_{l ;  I_3 I_4 \rightarrow \{\alpha\}}^{(0)}(s)]^\star~, \qquad \forall l \in \mathbb{N}~.
\end{equation}
Let us first check the unitarity of the purely bosonic one-loop result \eqref{55}. The non-vanishing tree-level coefficients are given by 
\begin{align} 
M^{(0)}_{0 ; I_1 I_2 \rightarrow I_3 I_4} &= \frac{s^2}{4} \Big( -\frac{1}{4} \delta^{I_1 I_2} \delta^{I_3 I_4} + \delta^{I_1 I_3} \delta^{I_2 I_4}  + \delta^{I_1 I_4} \delta^{I_2 I_3} \Big)~, \\
M^{(0)}_{1 ; I_1 I_2 \rightarrow I_3 I_4} &= \frac{s^2}{8} \left( \delta^{I_1 I_3} \delta^{I_2 I_4}  -  \delta^{I_1 I_4} \delta^{I_2 I_3} \right)~, \\
M^{(0)}_{2 ; I_1 I_2 \rightarrow I_3 I_4} &= \frac{s^2}{32} \delta^{I_1 I_2} \delta^{I_3 I_4}~,
\end{align}
while the imaginary part of the one-loop bosonic amplitude decomposes into
\begin{align} 
\Ima\, M^{(1)}_{0; I_1 I_2 \rightarrow I_3 I_4} &= \frac{1}{128} s^{7/2} \Big(\frac{\hat{D}-16}{32} \delta^{I_1 I_2} \delta^{I_3 I_4} + \delta^{I_1 I_3} \delta^{I_2 I_4} + \delta^{I_1 I_4} \delta^{I_2 I_3} \Big)~, \\
\Ima\, M^{(1)}_{1; I_1 I_2 \rightarrow I_3 I_4} &= \frac{1}{512} s^{7/2} \left(\delta^{I_1 I_3} \delta^{I_2 I_4} -\delta^{I_1 I_4} \delta^{I_2 I_3} \right)~, \\ 
\Ima\, M^{(1)}_{2; I_1 I_2 \rightarrow I_3 I_4} &= \frac{1}{256 \cdot 64} s^{7/2} \hat{D}\delta^{I_1 I_2} \delta^{I_3 I_4} ~.
\end{align}
One can then check that the unitarity relations
\begin{equation} 
\begin{aligned}
\Ima\, A_l^{(1)} &= \frac{1}{16 s^{1/2}} \left( \hat{D} |A_l^{(0)}|^2 + [A_l^{(0)}]^\star (B_l^{(0)} + C_l^{(0)} ) + A_l^{(0)} [B_l^{(0)} + C_l^{(0)} ]^\star  \right)~, \\
\Ima\, B_l^{(1)} &= \frac{1}{16 s^{1/2}} \left( |B_l^{(0)}|^2 + |C_l^{(0)}|^2 \right)~, \qquad \Ima\,  C_l^{(1)} = \frac{1}{16 s^{1/2}} \left( [B_l^{(0)}]^\star C_l^{(0)} + B_l^{(0)} [C_l^{(0)}]^\star\right)~,
\end{aligned}
\end{equation}
are indeed satisfied. Interestingly, while the amplitude \eqref{55} does not seem to single out a special value for $\hat{D}$, one sees from the above that the partial waves for the tree-level and one-loop amplitudes will have a similar structure in the special case of $\hat{D}=8$.

For the full amplitude, including the contribution coming from the fermions but restricting to the special $\hat{D}=8$ ($D=11$) case, the imaginary part of the one-loop amplitude decomposes into
\begin{align}
\Ima\, M^{(1)}_{0; I_1 I_2 \rightarrow I_3 I_4} &= \frac{1}{128} s^{7/2} \Big(-\frac{1}{4} \delta^{I_1 I_2} \delta^{I_3 I_4} + \delta^{I_1 I_3} \delta^{I_2 I_4} + \delta^{I_1 I_4} \delta^{I_2 I_3} \Big)~, \\
\Ima\, M^{(1)}_{1; I_1 I_2 \rightarrow I_3 I_4} &= \frac{1}{256} s^{7/2} \left(\delta^{I_1 I_3} \delta^{I_2 I_4} -\delta^{I_1 I_4} \delta^{I_2 I_3} \right)~, \\ 
\Ima\, M^{(1)}_{2; I_1 I_2 \rightarrow I_3 I_4} &= \frac{1}{256 \cdot 4} s^{7/2} \left(\delta^{I_1 I_2} \delta^{I_3 I_4} \right)~.
\end{align}
This is then in agreement with the additional contribution coming from the propagation of intermediate fermions,
\begin{equation} \begin{aligned}
-\int \frac{d \phi}{2 \pi}\Tr[\mathcal V_{I_1 I_2} \mathcal P_- \Gamma^\mu q_\mu \mathcal V_{I_3 I_4} \mathcal P_- \Gamma^\nu p_\nu] = & \frac{1}{32} s^{4} \cos \theta (\delta^{I_1 I_3} \delta^{I_2 I_4} - \delta^{I_1 I_4} \delta^{I_2 I_3}) +\frac{s^{4}}{128} \cos 2 \theta \delta^{I_1 I_2} \delta^{I_3 I_4}~,
\end{aligned}
\end{equation}
where
\begin{equation}
p_\mu = \frac{1}{2} s^{1/2} (1,\cos \varphi_{iv}, \sin \varphi_{iv})~, \qquad q= p-p_1-p_2~.
\end{equation}

\section{Tadpole diagram contributions}
\la{C}

Here we compute the contribution of the tadpole diagrams like in Fig. \ref{f3}. 
The relevant sextic interaction term $\mathcal L_6$ in~\rf{23} 
arises from the expansion of $S_1$ in~\rf{13} only.
After  $\kappa$-symmetry gauge fixing~\rf{28} and  rescaling of  the fermions as in~\rf{30}  we get 
\begin{align}
&\la{c1}\mathcal L_6 = \tfrac{1}{48} h_1^3 - \tfrac{1}{8} h_1^2 h_2 + \tfrac{1}{6} h_3~, \qquad\qquad 
 h_k = \Tr[(h_{\mu \nu})^k]~, \\
 &\la{c2}  h_{\mu \nu} = \partial_\mu X^I \partial_\nu X^I - \tfrac{i}{2} \bar{\theta} \mathcal P_-(\Gamma_\mu \partial_\nu + \Gamma_\nu \partial_\mu) \theta~.
\end{align}
The  6-boson interaction vertex  then reads (with all momenta assumed to be  incoming)
\begin{equation}
\begin{aligned}
&\mathcal V_{I_1,I_2,I_3,I_4,I_5,I_6}[p_1,p_2,p_3,p_4,p_5,p_6] = \delta^{I_1 I_2} \delta^{I_3 I_4} \delta^{I_5 I_6}  \\
\ \ \  & \times  \Big[- (p_1 \cdot p_2) (p_3 \cdot p_4) (p_5 \cdot p_6) + (p_1 \cdot p_2) (p_3 \cdot p_5) (p_4 \cdot p_6) +(p_1 \cdot p_2) (p_3 \cdot p_6) (p_4 \cdot p_5) \\
&\ \ \  +(p_1 \cdot p_3) (p_2 \cdot p_4) (p_5 \cdot p_6) - (p_1 \cdot p_3) (p_2 \cdot p_5) (p_4 \cdot p_6) - (p_1 \cdot p_3)(p_2 \cdot p_6)(p_4 \cdot p_5) \\
&\ \ \ +(p_1 \cdot p_4) (p_2 \cdot p_3) (p_5 \cdot p_6) - (p_1 \cdot p_4) (p_2 \cdot p_5) (p_3 \cdot p_6) - (p_1 \cdot p_4)(p_2 \cdot p_6)(p_3 \cdot p_5) \\
&\ \ \ +(p_1 \cdot p_5) (p_2 \cdot p_6) (p_3 \cdot p_4) - (p_1 \cdot p_5) (p_2 \cdot p_4)(p_3 \cdot p_6) - (p_1 \cdot p_5) (p_2 \cdot p_3) (p_4 \cdot p_6) \\
&\ \ \ +(p_1 \cdot p_6) (p_2 \cdot p_5) (p_3 \cdot p_4) - (p_1 \cdot p_6) (p_2 \cdot p_4)(p_3 \cdot p_5) - (p_1 \cdot p_6) (p_2 \cdot p_3) (p_4 \cdot p_5)  \Big] \\
&  + \text{permutations  of} \ (1,2,3,4,5,6)~.
\end{aligned}
\end{equation}
The  vertex between 4 bosons and 2 fermions is (all particles incoming, antiparticles outgoing)
\begin{equation}
\begin{aligned}
&(\mathcal V_{I_1,I_2,I_3,I_4})^{\alpha_5}_{\alpha_6}[p_1,p_2,p_3,p_4,p_5,p_6] = \tfrac{1}{2} \delta^{I_1 I_2} \delta^{I_3 I_4} \mathcal P_- \\
& \times  \Big[-(p_1 \cdot p_2) (p_3 \cdot p_4) (\Gamma \cdot p_6) + (p_1 \cdot p_2) (p_3 \cdot \Gamma) (p_4 \cdot p_6) + (p_1 \cdot p_2)(p_3 \cdot p_6)(p_4 \cdot \Gamma) \\
&\quad + (p_1 \cdot p_3) (p_2 \cdot p_4) (\Gamma \cdot p_6) - (p_1 \cdot p_3) (p_2 \cdot \Gamma) (p_4 \cdot p_6) - (p_1 \cdot p_3)(p_2 \cdot p_6)(p_4 \cdot \Gamma) \\
&\quad +(p_1 \cdot p_4) (p_2 \cdot p_3) (\Gamma\cdot p_6) - (p_1 \cdot p_4) (p_2 \cdot \Gamma) (p_3 \cdot p_6) - (p_1 \cdot p_4)(p_2 \cdot p_6)(p_3 \cdot \Gamma) \\
&\quad +(p_1 \cdot \Gamma) (p_2 \cdot p_6) (p_3 \cdot p_4) - (p_1 \cdot \Gamma) (p_2 \cdot p_4)(p_3 \cdot p_6) - (p_1 \cdot \Gamma) (p_2 \cdot p_3) (p_4 \cdot p_6) \\
&\quad +(p_1 \cdot p_6) (p_2 \cdot \Gamma) (p_3 \cdot p_4) - (p_1 \cdot p_6) (p_2 \cdot p_4)(p_3 \cdot \Gamma) - (p_1 \cdot p_6) (p_2 \cdot p_3) (p_4 \cdot \Gamma) 
\Big] \\
& + \text{permutations of } \ (1,2,3,4)~.
\end{aligned}
\end{equation}
Note that the  fermionic amplitude can be obtained from the bosonic one through the replacement 
$ \delta^{I_5 I_6} p_{5,\mu}p_{6,\nu}$ $\rightarrow$ $\tfrac{1}{2} \mathcal P_- \Gamma_\mu p_{6,\nu}~$
and setting to zero all terms without the $\delta^{I_5 I_6}$ index structure. A similar statement holds for the 4-point vertex contribution coming from $S_1$ only in~\rf{33} and~\rf{34}.

The  tadpole contribution of the  bosonic  loop   is represented  by the following integral 
\begin{equation}
\mathcal M_{B,\tad}  = \frac{v_{_B}}{i} \int \frac{d^d l}{(2 \pi)^d} \frac{N_B}{l^2- i \varepsilon}~, \qquad N_B = \sum_{I_l}\mathcal V_{I_1,I_2,I_3,I_4,I_l,I_l}[p_1,p_2,-p_3,-p_4,l,-l]~,  \la{c5}
\end{equation}
with $v_{_B}=-\ha$  being  the normalization constant   and  
 \begin{align}
&N_B=  N_2 l^2 + N_{\mu \nu} l^\mu l^\nu~, \ \qquad N_{2} = \tfrac{\hD-4}{2} \left[ t u \, \delta^{I_1 I_2} \delta^{I_3 I_4} + s u \,\delta^{I_1 I_3} \delta^{I_2 I_4}+ s t\, \delta^{I_1 I_4} \delta^{I_2 I_3}\right]~,\la{c6} \\
&N_{\mu \nu} = \left[(\hD-2) s (p_{1,\mu} p_{2,\nu}+p_{3,\mu} p_{4,\nu}) + \hD t (p_{1,\mu} p_{3,\nu}+p_{2,\mu} p_{4,\nu})+ \hD u (p_{1,\mu} p_{4,\nu}+p_{2,\mu} p_{3,\nu})\right]\delta^{I_1 I_2} \delta^{I_3 I_4} \no\\
&\quad  + \left[-(\hD-2) t (p_{1,\mu} p_{3,\nu}+p_{2,\mu} p_{4,\nu}) - \hD s (p_{1,\mu} p_{2,\nu}+p_{3,\mu} p_{4,\nu})+ \hD u (p_{1,\mu} p_{4,\nu}+p_{2,\mu} p_{3,\nu})\right]\delta^{I_1 I_3} \delta^{I_2 I_4} \no\\
&\quad + \left[-(\hD-2) u (p_{1,\mu} p_{4,\nu}+p_{2,\mu} p_{3,\nu}) + \hD t (p_{1,\mu} p_{3,\nu}+p_{2,\mu} p_{4,\nu})- \hD s (p_{1,\mu} p_{2,\nu}+p_{3,\mu} p_{4,\nu})\right]\delta^{I_1 I_4} \delta^{I_2 I_3}~,\no \\
& 
 \eta^{\mu \nu} N_{\mu \nu} = 2(s^2 - \hD t u ) \delta^{I_1 I_2} \delta^{I_3 I_4} +  2(t^2 - \hD s u) \delta^{I_1 I_3} \delta^{I_2 I_4} +  2(u^2 - \hD s t ) \delta^{I_1 I_4} \delta^{I_2 I_3}~.
\end{align}
The contribution from the fermionic loop is
\begin{equation}
\mathcal M_{F,\tad} = \frac{v_{_F}}{i} \int \frac{d^d l}{(2 \pi)^d} \frac{N_F}{l^2- i \varepsilon}~, \qquad N_F = \Tr \left[ \mathcal V_{I_1,I_2,I_3,I_4}[p_1,p_2,-p_3,-p_4,l,l] \mathcal P_-\Gamma^\mu l_\mu\right] ~,  \la{c7}
\end{equation}
with $v_{_F}=-1 $   and 
\begin{equation} \begin{aligned}
&N_F = N_2 l^2 + N_{\mu \nu} l^\mu l^\nu~, \qquad N_{2} = -\tfrac{1}{8}\nf \left( t u \, \delta^{I_1 I_2} \delta^{I_3 I_4} + s u \,\delta^{I_1 I_3} \delta^{I_2 I_4}+ s t\, \delta^{I_1 I_4} \delta^{I_2 I_3}\right)~, \\
&N_{\mu \nu} = -\tfrac{1}{4}\nf\Big\{ \left[s (p_{1,\mu} p_{2,\nu}+p_{3,\mu} p_{4,\nu}) + t (p_{1,\mu} p_{3,\nu}+p_{2,\mu} p_{4,\nu})+  u (p_{1,\mu} p_{4,\nu}+p_{2,\mu} p_{3,\nu}) \right]\delta^{I_1 I_2} \delta^{I_3 I_4} \\
& + \left[ -t (p_{1,\mu} p_{3,\nu}+p_{2,\mu} p_{4,\nu}) - s (p_{1,\mu} p_{2,\nu}+p_{3,\mu} p_{4,\nu})+ u (p_{1,\mu} p_{4,\nu}+p_{2,\mu} p_{3,\nu})\right]\delta^{I_1 I_3} \delta^{I_2 I_4} \\
& + \left[- u (p_{1,\mu} p_{4,\nu}+p_{2,\mu} p_{3,\nu}) + t (p_{1,\mu} p_{3,\nu}+p_{2,\mu} p_{4,\nu})- s (p_{1,\mu} p_{2,\nu}+p_{3,\mu} p_{4,\nu})\right]\delta^{I_1 I_4} \delta^{I_2 I_3}\Big\}~, \\
&
 \eta^{\mu \nu} N_{\mu \nu} = \tfrac{1}{2}\nf\big(  t u  \delta^{I_1 I_2} \delta^{I_3 I_4} +  s u \delta^{I_1 I_3} \delta^{I_2 I_4} +  s t \delta^{I_1 I_4} \delta^{I_2 I_3}\big) ~.
\end{aligned}
\end{equation}
Using that   $\int d^d l \,  f(l^2)\,  l^\mu l^\nu = \frac{1}{d} \eta^{\mu \nu} \int d^d l\,   f(l^2)\,  l^2 $, we get  for $d=3$ 
\begin{align}
 \mathcal M_\tad =& \mathcal M_{B,\tad} + \mathcal M_{F,\tad} =  J \Big\{\, 
2 v_{_B}  \left( s^2\, \delta^{I_1 I_2} \delta^{I_3 I_4} + t^2 \,\delta^{I_1 I_3} \delta^{I_2 I_4}+ u^2\, \delta^{I_1 I_4} \delta^{I_2 I_3}\right) \\
&+\big[  - \half v_{_B} ( \hD + 12 )  + \tfrac{1}{8} v_{_F} \nf  \big] \left( t u \, \delta^{I_1 I_2} \delta^{I_3 I_4} + s u \,\delta^{I_1 I_3} \delta^{I_2 I_4}+ s t\, \delta^{I_1 I_4} \delta^{I_2 I_3}\right)\Big\}~, \no \\
&\qquad \qquad J= d^{-1}\int \frac{d^d l}{(2\pi)^d} \frac{l^2}{l^2-i \varepsilon}\ . 
\end{align}
For   $v_{_F}=2 v_{_B} =-1 , \ \hD=\ha \nf=8$  we get explicitly 
\be
\mathcal M _\tad  =  J \Big[ (4 tu -s^2) \delta^{I_1 I_2} \delta^{I_3 I_4}  +  (4 su -t^2) \,\delta^{I_1 I_3} \delta^{I_2 I_4}
+ (4st - u^2\, \delta^{I_1 I_4}) \delta^{I_2 I_3} \Big] \ . 
\ee
In $d=3$  the factor $J$ is a cubic or $\delta^{(3)}(0)$ divergence    that may  get 
 cancelled against a  local path integral measure factor or automatically set to 0 in dimensional  regularization. 

In the case of the  cylindrical $\mathbb R^{1,1} \times S^1$ supermembrane, 
with massless external bosons but  all  modes propagating in the loop in the tadpole diagram
to find the contribution from level $n$ modes
 one needs to  replace  $l^2 \rightarrow l^2 + n^2$ in $N_{B}$  and $N_{F}$ in~\rf{c5} and~\rf{c7} 
 as well as in the propagator. 
 Evaluating the integral over the loop momentum $l$ in dimensional regularisation 
 with $d=2-2 \epsilon$ then leads to the   tadpole  contribution 
\begin{align}
&\hat{\mathcal M}_{\tad,n} = \tfrac{1}{4 \pi}\Big[  n^2\left( \tfrac{1}{\epsilon} - \gamma + \ln 4 \pi  + 1\right)  - n^2 \ln n^2 \Big] \Big\{\big[
- v_{_B} s^2 +( v_{_B} \hD - \tfrac{1}{4}  v_{_F} \nf) t u\big] 
 \delta^{I_1 I_2} \delta^{I_3 I_4}\no  \\
&+\big[- v_{_B} t^2 + (v_{_B} \hD - \tfrac{1}{4} v_{_F}\nf) s u \big]
 \delta^{I_1 I_3} \delta^{I_2 I_4} +\big[- v_{_B} u^2 + (v_{_B} \hD  - \tfrac{1}{4}  v_{_F}   \nf)\, s t \big]
 \delta^{I_1 I_4} \delta^{I_2 I_3} \Big\}~,\la{c8}
\end{align}
that should be added to the bubble diagram contribution in~\rf{67}. 
We observe that while the $n^2$ terms 
in~\rf{c8}  give zero  after the summation over $n$,  
setting $v_{_B} = -\tfrac{1}{2}$ and $v_{_F}=-1$   and  $\hD = \tfrac{\nf}{2}$ 
 the $n^2 \ln n^2$ terms cancel against  similar terms in~\rf{67}. 
 Equivalently,  after the summation over $n$, the tadpole contribution cancels the $\zeta(3)$ term in~\rf{71} 
 leading to~\rf{711}.

\iffa 
For   the $n^2 \ln n^2$ contribution from
 the tadpole diagram cancels similar terms in the one-loop bubble amplitude, as claimed in~\rf{711} in the main text. It also cancels the $n^2$ contribution to the divergent amplitude. This is true for all $\hD$, as long as the supersymmetric constraint  is satisfied.
\fi

\iffa 
\section{From 10D to 2d fermions}
The Green-Schwarz action before integrating out the auxiliary metric reads
\begin{equation}
S = S_1 + S_2~, \qquad  S_1 = -T_1 \int d^2 \sigma \frac{1}{2} \sqrt{-\det g_{\mu \nu}} g^{\mu \nu} \Pi_{\mu}^N \Pi_{\nu}^N \eta_{MN}~,
\end{equation}
with $S_2$ the Wess-Zumino term left unchanged. To quadratic order in the fermions, this is
\begin{equation}
S_1 = - T_1 \int d^2 \sigma \frac{1}{2} \sqrt{-\det g_{\mu \nu}}  \left( g^{\mu \nu} \partial_\mu X^M \partial_\nu X_M  - 2 i \bar{\theta}  \Gamma^\nu \partial_\nu \theta \right)~, \qquad \Gamma^\nu = g^{\mu \nu}  \partial_\mu X^M \Gamma_M~.
\end{equation}
Here $\theta$ are Majorana spinors in $D$-dimensional target-space, and scalars on the two-dimensional worldsheet.
On the other hand, the action for 2d Dirac fermions $\psi$ reads
\begin{equation}
\tilde{S}_1 = - T_1 \int d^2 \sigma \frac{1}{2} \sqrt{-\det g_{\mu \nu}} \left( g^{\mu \nu}\partial_\mu X^M \partial_\nu X_M - i \bar{\psi}  e_a^\mu \tau^a \partial_\mu \psi \right)~,
\end{equation}
where $\tau^a$ are $2 \times 2$ Dirac matrices.
If we take
\begin{equation}
\Gamma_a = \tau_a \otimes 1_{16}~, \qquad \Gamma_I = \tau^\star \otimes \gamma_I
\end{equation}
\fi

\small 
\bibliographystyle{JHEP-v2.9}
\small
\bibliography{biblio2.bib}

\end{document}